\documentclass[twoside,a4paper,11pt]{article}
\usepackage{longtable} 
\usepackage{blindtext}
\usepackage{amsmath}
\usepackage{booktabs}
\usepackage{multirow}
\usepackage{tikz}
\usepackage{bm}       

\usetikzlibrary{arrows.meta, positioning, shapes.geometric, fit, backgrounds, decorations.pathreplacing}
\usepackage[table]{xcolor}
\definecolor{tcol}{HTML}{7570B3}
\definecolor{ecol}{HTML}{FF7F0E}

\usepackage{subcaption}
\usepackage{xcolor}

\usepackage{tabularx}

\usepackage{irrj2e}
\usepackage{algorithm}      
\usepackage{algpseudocode}

\usepackage{lastpage}

\irrjheading{1}{2025}{1-\pageref{LastPage}}{2/25}{-}{10.54195/irrj.00000}

\ShortHeadings{Scalable Music Cover Retrieval Using Lyrics-Aligned Audio Embeddings}{J. Affolter, B. Martin, G. Meseguer-Brocal and E. V. Epure}
\firstpageno{1}

\begin{document}

\title{Scalable Music Cover Retrieval using Lyrics-Aligned Audio Embeddings:
An Extended Study}

\author{\name Joanne Affolter \email jaffolter@deezer.com \\
       \addr Deezer Research - 24 Rue de Calais, 75009 Paris
       \AND
       \name Benjamin Martin \email research@deezer.com \\
       \addr Deezer Research - 24 Rue de Calais, 75009 Paris
       \AND
       \name Gabriel Meseguer-Brocal \email research@deezer.com \\
       \addr Deezer Research - 24 Rue de Calais, 75009 Paris
       \AND
       \name Elena V. Epure \email research@deezer.com \\
       \addr Deezer Research - 24 Rue de Calais, 75009 Paris \\ 
       \addr Idiap Research Institute - 19 Rue Marconi, 1920 Martigny 
       \AND
       \name Frédéric Kaplan \email frederic.kaplan@epfl.ch \\
        \addr EPFL DHLAB - Rte Cantonale, 1015 Lausanne
       }

\editor{My editor}

\maketitle

\begin{abstract}
Music Cover Retrieval, also known as Version Identification, aims to recognize distinct renditions of the same underlying musical work, a task central to catalog management, copyright enforcement, and music retrieval.
State-of-the-art approaches have largely focused on harmonic and melodic features, employing increasingly complex audio pipelines designed to be invariant to musical attributes that often vary widely across covers. While effective, these methods demand substantial training time and computational resources.
By contrast, lyrics constitute a strong invariant across covers, though their use has been limited by the difficulty of extracting them accurately and efficiently from polyphonic audio. 
Early methods relied on simple frameworks that limited downstream performance, while more recent systems deliver stronger results but require large models integrated within complex multimodal architectures.
We introduce LIVI (Lyrics-Informed Version Identification), an approach that seeks to balance retrieval accuracy with computational efficiency.
First, LIVI leverages supervision from state-of-the-art transcription and text embedding models during training to achieve retrieval accuracy on par with—or superior to—harmonic-based systems. Second, LIVI remains lightweight and efficient by removing the transcription step at inference, challenging the dominance of complexity-heavy pipelines.
\end{abstract}


\section{Introduction}
\label{sec:introduction}

In information retrieval, tasks such as Near-Duplicate Detection~\citep{rodier-carter-2020-online, tumre-etal-2025-improved} and Entity Resolution~\citep{zhao-etal-2020-graph} aim to identify and link semantically equivalent entities. In the music information retrieval domain, an analogous task is Music Cover Retrieval—also known as Version Identification or Cover Detection—where the goal is to recognize distinct renditions of the same underlying composition~\citep{yesiler_thesis}.
Robust systems are critical for catalog management, copyright enforcement, cross-platform track linking, and music retrieval~\citep{yesiler_thesis}.
 
Defining similarity between covers, however, is challenging: models must account for wide variations in tempo, timbre (instrumentation and vocal characteristics), pitch (transposition to different keys or modal changes), structure (addition, removal, or reordering of sections), lyrical content (translation or rewriting), recording conditions, and performance interpretation~\citep{yesiler_thesis}. State-of-the-art approaches have largely focused on harmonic and melodic features, often relying on complex pipelines that aim to achieve invariance to these attributes~\citep{du2021bytecovercoversongidentification, hu22f_lyracnet, liu2023Coverhunter}. While effective, such models demand significant training time and computational resources, and their growing reliance on deep or multimodal architectures further limits scalability and reproducibility~\citep{abrassart2022musicalversionsdontshare}.

Unlike melodic or harmonic features, which may vary widely across renditions, prior work has shown that lyrics constitute a strong invariant~\citep{abrassart2022musicalversionsdontshare, correya2018largescale, du2024xcover, mancini2025leveragingwhisperembeddingsaudiobased, vaglio2021words}: (i) they are typically preserved across most renditions, with notable exceptions such as parody, (ii) they largely retain their semantic content even under translation or minor rewrites, and (iii) they provide decisive cues for distinguishing works that share similar harmonic or melodic profiles. A well-known illustration is Jimi Hendrix's cover of ``All Along the Watchtower'' from Bob Dylan, where harmony and melody diverge strongly from the original, yet the lyrics remain largely intact.
Despite this, their potential has been underexplored, mainly due to two obstacles: the availability of editorial lyrics at scale, often restricted by third-party licensing, and the challenge of extracting lyrics from polyphonic audio~\citep{yesiler_thesis}. While early works that use lyrics ~\citep{abrassart2022musicalversionsdontshare, vaglio2021words} rely on relatively simple approaches to derive lyric representations, resulting in limited downstream performance, more recent work~\citep{du2024xcover} achieve stronger results but integrate transcription into a complex multimodal architecture, increasing model size and computational cost.

Our work builds on the hypothesis that songs with semantically similar lyrics are likely to be covers~\citep{correya2018largescale}. To test this idea, we first construct a pipeline that represents songs in a lyric-informed embedding space, obtained by applying an Automatic Speech Recognition (ASR) system followed by a multilingual text encoder. This design is motivated by two factors: (i) clean editorial lyrics are rarely available at scale, making transcription a necessary step, and (ii) modern multilingual text encoders, pretrained for semantic similarity, provide a powerful and readily applicable representation space. While this pipeline achieves strong performance, its reliance on full transcription makes it computationally costly. Consequently, we introduce LIVI (Lyrics-Informed Version Identification), a model that learns to project latent audio representations directly into the lyric embedding space defined by the pipeline, removing the transcription step and thereby reducing inference cost while preserving retrieval accuracy.

Despite its relative simplicity and the absence of explicit fine-tuning for the downstream task, LIVI achieves performance on par with—or superior to—state-of-the-art systems. LIVI delivers an efficient, reproducible\footnote{\scriptsize Code available at \url{https://github.com/deezer/LIVI-Lyrics-Informed-Version-Identification}}, and domain-grounded alternative, challenging the dominance of complexity-heavy multimodal systems. By design, our method applies only to tracks with sufficient vocal content, with a preprocessing stage used to exclude those lacking it. While this restriction narrows the scope of applicability, its practical impact is limited given the predominance of vocal music in mainstream repertoires and its central role in industrial applications~\citep{demetriou-etal-2018-vocals}. Moreover, a lyrics-informed approach such as LIVI could naturally be complemented by harmonic features, as in~\citet{vaglio2021words}, forming part of a broader system that integrates both textual and musical cues.

This article extends LIVI~\citep{livi_ecir} along two main axes. 
First, we include a new component that targets ASR hallucinations, a phenomenon known to occur on non-vocal audio~\citep{calm_whisper, careless_whisper, hallucinations, Bara_ski_2025}. Since LIVI is trained to align with lyrics embeddings derived from ASR transcriptions, such hallucinations pollute the target embedding space and degrade retrieval for the affected tracks. \citet{livi_ecir} use a proprietary vocal-detection model to discard purely instrumental audio, but some hallucinations still persist for a subset of vocal tracks, notably in genres such as opera and scat-style jazz where vocals are too atypical or sparse to transcribe reliably. We address this by replacing the proprietary model with an open, lightweight classifier trained to identify both instrumental and hallucination-prone chunks. The classifier exploits the fact that ASR hallucinations are drawn from a small, predictable vocabulary of training-data artifacts~\citep{Bara_ski_2025}, allowing the same ASR encoder features used by LIVI to also predict whether a chunk is likely to hallucinate. 
Second, we include a new reranking component that recovers local information otherwise lost in track-level representations. Because covers frequently reorder, omit, or rewrite sections~\citep{yesiler_thesis}, two versions may align on a distinctive verse or chorus while diverging globally; averaging all chunks into a single vector dilutes precisely these local matches. Motivated by this, we introduce a two-stage retrieve-then-rerank design that pairs fast global-embedding search for candidate generation with a parameter-free MaxSim late-interaction step~\citep{khattab2020colbert} over local embeddings. We show that this recovers versions sharing strong local matches, while improving Music Cover Retrieval performance across all benchmarks considered and preserving the efficiency that motivates LIVI's design.

The remainder of this article is organized as follows. Section~\ref{sec:related} reviews related work. Section~\ref{sec:methodo} formalizes the retrieval problem and presents the full framework. Section~\ref{sec:livi} details the implementation. Section~\ref{sec:setup} describes the experimental setup, Section~\ref{sec:results} reports quantitative results, and Section~\ref{sec:qualitative} provides a qualitative analysis of the learned embedding space. Section~\ref{sec:conclusion} concludes.

\section{Related Works}
\label{sec:related}
 
\subsection{Version Identification}
Research on Music Cover Retrieval has undergone several stages of development over the past two decades. Early systems based on hand-crafted features achieved encouraging results on small benchmarks, but failed to scale due to their reliance on costly alignment techniques such as dynamic time warping (DTW)~\citep{Müller2007, yesiler_thesis}. The advent of deep learning marked a turning point, enabling data-driven feature learning from harmonic and melodic representations such as predominant melody, pitch class profiles (PCP), or the constant-Q transform (CQT)~\citep{yesiler2020accuratescalableversionidentification}. Since then, progress has followed two main directions: one line of work pushes towards increasingly deep architectures such as ResNet~\citep{araz2024discogsvinet_mirex, du2024xcover, bytecover2, du2023bytecover3accuratecoversong, du2021bytecovercoversongidentification, hu22f_lyracnet, liu2023Coverhunter, serrà2025supervisedcontrastivelearningweaklylabeled, yu_2019TPP, 9053839}, while another emphasizes musically informed inductive biases to achieve invariance to transformations such as transposition or structural changes~\citep{abrassart2022musicalversionsdontshare, doras2019coverdetectionusingdominant, Doras2020CombiningMF, yesiler2020accuratescalableversionidentification, yesiler2020morefasterbettermusic}. Beyond these, multimodal methods that integrate complementary features have demonstrated clear advantages over unimodal models~\citep{abrassart2022musicalversionsdontshare, Doras2020CombiningMF, du2024xcover, vaglio2021words, mancini2025leveragingwhisperembeddingsaudiobased}. Table~\ref{tab:cover-models} summarizes the main characteristics of these approaches.
 
Although lyrics offer a strong discriminative signal for version identification~\citep{abrassart2022musicalversionsdontshare, vaglio2021words, du2024xcover, mancini2025leveragingwhisperembeddingsaudiobased}, their integration into Music Cover Retrieval systems is relatively recent, primarily due to the absence of large-scale datasets with clean, time-aligned lyrics and the difficulty of transcribing lyrics from polyphonic audio~\citep{yesiler_thesis}. In early attempts, \citet{vaglio2021words} adopted a Singing Voice Recognition (SVR) framework combining a TDNN-based acoustic model~\citep{fan2020deeptimedelayneural} with a language model to decode phoneme sequences compared via string matching. \citet{abrassart2022musicalversionsdontshare} proposed a lightweight Automatic Lyrics Recognition (ALR) model that produced character posteriorgrams subsequently processed by a second, independently trained model fine-tuned for the retrieval task. Although these approaches marked important first steps, they relied on acoustic models to represent lyrics, yielding limited retrieval performance. Moreover, transcription models were trained from scratch on English-only datasets \citep{https://doi.org/10.5281/zenodo.1492443}, restricting multilingual generalization and adding the overhead of training a separate retrieval model.
 
Recent advances in lyrics-based cover detection include dedicated datasets, such as LyricsCovers~2.0~\citep{lyriccovers20}, and new methods~\citep{langindep_phoneme, du2024xcover, balluff2024innovationscoversongdetection, mancini2025leveragingwhisperembeddingsaudiobased}. Among these, \citet{du2024xcover} achieved stronger results by adapting Whisper~\citep{radford2022robustspeechrecognitionlargescale} with lightweight prefix- and suffix-tuning, enabling faster inference while keeping most parameters frozen. However, the approach remains integrated into a complex multimodal pipeline, increasing model size and computational cost, and making it difficult to isolate the impact of lyrics. Reproducibility is further constrained by the lack of open-source implementations and technical details. More recently,~\citet{mancini2025leveragingwhisperembeddingsaudiobased} introduced WEALY, a pipeline that extracts decoder hidden states from the full frozen Whisper model and trains a transformer encoder via contrastive learning on the version identification task, bypassing explicit transcription at inference. While this confirms the discriminative value of Whisper's internal representations for Music Cover Retrieval, WEALY still requires running the computationally heavy autoregressive decoding process, and performance remains below state-of-the-art instrumental baselines.
In this work, we also leverage lyrics as a discriminative signal for Music Cover Retrieval, but our design is driven primarily by efficiency.  Unlike prior lyrics-based approaches, which integrate lyrics within larger multimodal systems~\citep{abrassart2022musicalversionsdontshare, vaglio2021words, du2024xcover}, we rely solely on lyrics-derived features. Furthermore, in contrast to~\citet{mancini2025leveragingwhisperembeddingsaudiobased} and ~\citet{du2024xcover}, our model removes the dependency on autoregressive decoding and achieves performance on par with audio-based approaches through a lightweight system.

\begin{table*}[t]
\centering
\resizebox{\textwidth}{!}{%
\begin{tabular}{@{}l l  l l  c c@{}}
\toprule
Name &  Modality & Features & Architecture &  \begin{tabular}[c]{@{}c@{}}Segment\\ based\end{tabular} & \begin{tabular}[c]{@{}c@{}}Open\\ source\end{tabular} \\
\midrule
CQTNet       \citep{yu_2019TPP}      & Audio          & CQT                                       & ConvNet       & No  & Yes \\
Vaglio       \citep{vaglio2021words}   & Lyrics + Audio & MFCC + PCP                                          & TDNN + LM        & No  & No  \\
LyraC-Net    \citep{hu22f_lyracnet}    & Audio          & CQT                          & ResNet  & No  & No  \\
Me+Ha+Ly     \citep{abrassart2022musicalversionsdontshare}& Lyrics + Audio & CQT + PCP + MS              & ConvNet        & No  & No  \\
ByteCover2   \citep{bytecover2}  & Audio          & CQT                               & ResNet   & No  & No  \\
ByteCover3   \citep{du2023bytecover3accuratecoversong}  & Audio          & CQT                         & ResNet  & Yes & No  \\
CoverHunter  \citep{liu2023Coverhunter}& Audio          & CQT                               & Conformer       & Yes & Yes \\
DVI-Net      \citep{araz2024discogsvinet_mirex} & Audio          & CQT                                      & ConvNet         & No  & Yes \\
X-Cover      \citep{du2024xcover}      & Lyrics + Audio & CQT + MS          & ResNet        & Yes  & No  \\
CLEWS        \citep{serrà2025supervisedcontrastivelearningweaklylabeled}    & Audio          & CQT                                         & ResNet      & Yes & Yes \\
WEALY        \citep{mancini2025leveragingwhisperembeddingsaudiobased}  & Lyrics     & MS                      & Transformer    & Yes  & Yes \\
\midrule
LIVI (Proposed) & Lyrics & MS & Transformer  & Yes & Yes \\
\bottomrule
\end{tabular}%
}
\caption{Overview of Version Identification models. 
{\scriptsize CQT (Constant-Q Transform), MFCC (Mel-Frequency Cepstral Coefficients), PCP (Pitch Class Profile), MS (Mel-Spectrogram).}}
\label{tab:cover-models}
\end{table*}
 
\subsection{Automatic Lyrics Recognition}
Modern Automatic Speech Recognition (ASR) systems are based on end-to-end neural architectures that map raw audio directly to text, in contrast to older hybrid HMM-DNN pipelines~\citep{6681449}. Among these, Whisper~\citep{radford2022robustspeechrecognitionlargescale} has emerged as the de facto open-source standard. Trained on 680,000 hours of labeled audio-text pairs collected from the web, it achieves broad coverage across languages, accents, and acoustic conditions. As an encoder-decoder Transformer, Whisper encodes audio into latent representations that the decoder attends to via cross-attention to generate text autoregressively, guided by special tokens specifying the task, target language, and formatting. Beyond transcription, Whisper natively supports speech translation, voice activity detection, and language identification~\citep{radford2022robustspeechrecognitionlargescale}. Its robustness to noise and real-world variability has made it particularly suitable for the heterogeneous conditions of music audio~\citep{cífka2024lyricstranscriptionhumansreadabilityaware, zhuo2024lyricwhizrobustmultilingualzeroshot}, and it has been widely adopted for automatic lyric transcription~\citep{syed2025exploitingmusicsourceseparation}. State-of-the-art accuracy is achieved by proprietary systems such as AudioShake v3~\citep{cífka2024lyricstranscriptionhumansreadabilityaware}, but Whisper remains the strongest open-source alternative.
 
A practical limitation of speech-trained ASR systems such as Whisper is their tendency to produce hallucinations when applied to non-vocal audio~\citep{careless_whisper, Bara_ski_2025}. \citet{Bara_ski_2025} provide a systematic characterization of this phenomenon, showing that Whisper hallucinates on over 40\% of diverse non-speech audio inputs. 
More than half arise from a small, stable vocabulary of training-data artifacts, reflecting Whisper’s pretraining on video transcriptions (e.g hallucinations such as \textit{``thank you''} or \textit{``thanks for watching''}), making them structurally predictable. \citet{Bara_ski_2025} address this as a post-processing problem, constructing a bag of hallucinations (BoH) by transcribing non-vocal audio and retaining the most frequent outputs, then filtering them out of downstream transcriptions. We adopt this frequency-based construction on audio chunks identified as instrumental, retaining phrases above a validated threshold to build the BoH used in our pipeline (Section~\ref{sec:weak_labels}).
 
\subsection{Audio to Text Alignment}
Recent advances in audio--text modeling focus on learning aligned representations across modalities, typically through contrastive pretraining in a shared embedding space. Inspired by CLIP~\citep{radford2021learningtransferablevisualmodels}, CLAP-style models~\citep{elizalde2022claplearningaudioconcepts, guzhov2021audioclipextendingclipimage, huang2022mulanjointembeddingmusic, yuan2024tclaptemporalenhancedcontrastivelanguageaudio, 2506.11350} jointly train audio and text encoders to maximize similarity between paired data, achieving strong zero-shot performance in tagging, retrieval, and captioning. However, these methods typically rely on high-level textual descriptors rather than structured content such as lyrics.
 
Several works have instead explored aligning audio with lyrics, though with different objectives. \citet{Durand_2023} proposed a contrastive learning framework in which singing audio and lyric transcripts are encoded as sequences of frame- and token-level embeddings. A similarity matrix is then computed to determine the optimal alignment path, enabling precise word-level synchronization. \citet{yu2017deepcrossmodalcorrelationlearning} addressed cross-modal retrieval by training parallel audio and lyric encoders with a Deep Canonical Correlation Analysis (DCCA) loss. Their approach projects spectrogram-based audio features and text embeddings into a joint space, but the reliance on correlation objectives makes training computationally expensive. Similar to these, we also derive lyrics representations from audio, motivated by the strong invariance of lyrics across versions for Music Cover Retrieval. Our method departs from these approaches in two respects: it integrates lyric semantics directly into the audio representation without requiring costly alignment or correlation objectives, and it operates on 30-second chunks---a coarser granularity than previous frame- or token-level approaches, which we show remains effective at the song level when averaged into a single embedding.
 
\subsection{Late Interaction}

The interaction paradigm between query and document representations is a central design choice in retrieval, governing the trade-off between computational efficiency and ranking precision~\citep{karpukhin2020dpr, reimers2019sentencebertsentenceembeddingsusing, nogueira2019passage, khattab2020colbert}. Bi-encoders~\citep{karpukhin2020dpr, reimers2019sentencebertsentenceembeddingsusing} encode queries and documents independently into single dense vectors, enabling fast nearest-neighbour search, while cross-encoders~\citep{nogueira2019passage} jointly encode each query-document pair to produce a more precise relevance score, at a per-pair cost that scales linearly with the candidate set. Late interaction, introduced by ColBERT~\citep{khattab2020colbert}, offers a middle ground: queries and documents are encoded independently as sets of contextualised token embeddings and scored by a MaxSim operator that, for each query token, takes its maximum similarity to any document token and sums these over the query, recovering local relevance signals without the cost of joint encoding. 
 
The Music Cover Retrieval literature has developed analogous segment-level approaches, albeit without the formal late-interaction framework. \citet{liu2023Coverhunter} work with non-overlapping 45-second local embeddings and resolve track identity by closest-chunk matching; \citet{du2024xcover} aggregate local embeddings via a metric similar to MaxSim; and~\citet{serrà2025supervisedcontrastivelearningweaklylabeled} propose a new metric, BPWR-r, which matches non-overlapping segment pairs iteratively to better capture partial overlaps. 
We adopt a comparable strategy to \citet{du2024xcover}, employing a MaxSim operator, but to different effect: while \citet{du2024xcover} incorporate it into the training objective, as does \citet{serrà2025supervisedcontrastivelearningweaklylabeled} with their own BPWR-r metric, we reserve it for inference, applying it as a parameter-free step to rerank a candidate set retrieved via global embeddings. 
This design keeps training computationally lighter.

\section{Methodology}
\label{sec:methodo}

We first present our approach to \textit{train} LIVI (Section~\ref{sec:training}), which leverages the invariance of lyrics across renditions to identify covers (Figure~\ref{fig:overview_livi}). Our starting point is a lyrics-informed pipeline designed to maximize retrieval accuracy, without regard to efficiency. Audio is first transcribed using an encoder–decoder ASR model, and the resulting text is then embedded with a multilingual model fine-tuned for semantic similarity. This produces an embedding space in which semantically similar lyric cluster closely, while unrelated text lies further apart.
Its key drawback lies in the reliance on transcription, where the ASR autoregressive decoder introduces considerable computational overhead. To overcome this, LIVI discards the decoder and trains an audio encoder to map latent ASR states directly into the lyric-informed embedding space derived from the pipeline. This removes the need for full transcription while preserving retrieval accuracy, resulting in a more efficient and scalable solution.
We then present our \textit{inference} pipeline (Section~\ref{sec:inference}), which operates in two stages (Figure~\ref{fig:inference_pipeline}). First, a set of candidates is retrieved using track-level embeddings. Second, all candidates are reranked using a MaxSim operator over 30-second local embeddings~\citep{khattab2020colbert}, enabling the system to focus on local similarity where global comparisons fall short. This design targets robustness to structural changes between versions—such as the addition, removal, or reordering of sections—known to be characteristic of covers~\citep{yesiler_thesis}, a direction recent version identification systems have also pursued successfully~\citep{du2023bytecover3accuratecoversong, serrà2025supervisedcontrastivelearningweaklylabeled}.
Before detailing these contributions, we formalize the retrieval problem in the next section.

\subsection{Problem Formulation}
\label{sec:formulation}
 
We formulate Music Cover Retrieval as a similarity ranking problem over embeddings produced by an encoder $g$. Let $\mathcal{C}$ be a catalog of music tracks and $g:\mathcal{C} \to \mathbb{R}^d$ map each track $x \in \mathcal{C}$ to an embedding $\mathbf{e}_x \in \mathbb{R}^d$.
Given a query $q \in \mathcal{C}$ (i.e., a song), the system assigns to each $x \in \mathcal{C} \setminus \{q\}$ a cosine similarity score
$$s(q,x)=\mathrm{cos}(\mathbf{e}_q,\mathbf{e}_x)
=\frac{\mathbf{e}_q^\top \mathbf{e}_x}{\|\mathbf{e}_q\|_2\,\|\mathbf{e}_x\|_2}$$
and returns the catalog ordered in descending order by $s(q,\cdot)$. Let $\mathcal{V}(q)\subset\mathcal{C}$ denote the set of versions of $q$. The desired ranking property is
$$s(q,v^+)>s(q,v^-) \qquad
\forall v^+ \in \mathcal{V}(q),\;\;v^-\in \mathcal{C}\setminus\mathcal{V}(q)$$
Accordingly, the encoder must learn an embedding space in which versions are embedded more closely than non-versions.

\subsection{LIVI Training}
\label{sec:training}

\begin{figure}[b!]
\centering
\includegraphics[width=\linewidth]{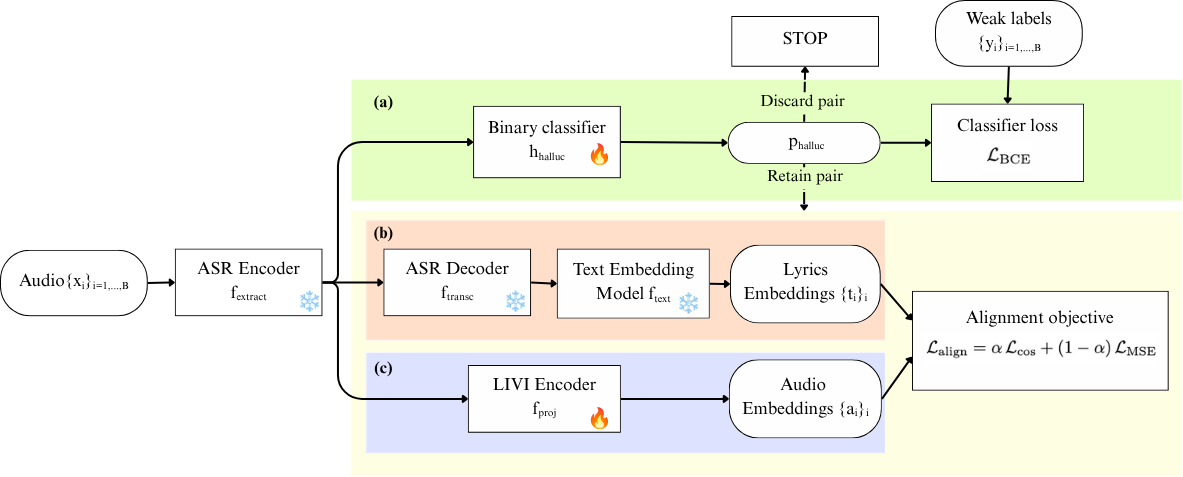}
\caption{\textbf{Overview of the LIVI framework.} A batch of 30\,s audio chunks $\{x_i\}$ is encoded once by the frozen ASR encoder $f_{\text{extract}}$, whose hidden states feed three branches sharing this backbone.\\ 
{\small 
(a) the binary classifier $h_{\text{halluc}}$ outputs a hallucination probability $p_{\text{halluc}}$: chunks with $p_{\text{halluc}} \geq \delta$ are discarded during training; \\
(b) the teacher path $g_{\text{text}} = f_{\text{text}} \circ f_{\text{transc}}$ completes transcription through the ASR decoder $f_{\text{transc}}$ and encodes the resulting transcript with a text encoder $f_{\text{text}} $; \\
(c) the LIVI encoder $f_{\text{proj}}$ projects the ASR encoder's latent representations into the lyrics-informed embedding space, yielding audio embeddings $a_i$; \\ 
The two models $h_{\text{halluc}}$ and $f_{\text{proj}}$ are trained independently: $f_{\text{proj}}$ by the alignment objective $\mathcal{L}_{\text{align}} = \alpha\,\mathcal{L}_{\text{cos}} + (1-\alpha)\,\mathcal{L}_{\text{MSE}}$, and $h_{\text{halluc}}$ by binary cross-entropy $\mathcal{L}_{\text{BCE}}$ against weak labels
$\{y_i\}$.} }
\label{fig:overview_livi}
\end{figure}

In multimodal representation learning, a natural approach to aligning audio and text representations  is contrastive learning~\citep{elizalde2022claplearningaudioconcepts, huang2022mulanjointembeddingmusic, wu2024largescalecontrastivelanguageaudiopretraining}, which jointly trains both modality encoders so that paired samples attract and unpaired ones repel in a shared space. 
Our method differs in that we keep the lyrics-informed embedding space fixed (Figure~\ref{fig:overview_livi}.b), exploiting its semantic structure (validated empirically on our downstream task in Section~\ref{sec:val-space}) as a supervision signal. We train only the audio encoder (Figure~\ref{fig:overview_livi}.c) to align with this space under a student-teacher scheme. 
Unlike contrastive objectives, our formulation avoids relying on in-batch negatives. At the granularity of 30-second chunks, a batch can contain two chunks from the same track, or chunks from different tracks that are semantically close; a contrastive objective would treat such pairs as negatives and push them apart, injecting noise into training. By instead treating the lyrics-informed space as fixed and transferring its known pairwise similarities onto the audio embedding space, our formulation provides more stable targets throughout training.

This lyrics-informed space, however, carries a vulnerability inherited from its construction process. Because transcription models are trained primarily on speech, they tend to hallucinate in non-vocal sections, generating spurious outputs despite the absence of linguistic content~\citep{calm_whisper, careless_whisper, hallucinations, Bara_ski_2025}. Music audio inevitably involves instrumental sections, and some tracks contain lyrics that are indistinct, sparse, or out-of-vocabulary (e.g jazz, opera), making them similarly prone to hallucination. Left unaddressed, these spurious transcriptions corrupt the lyrics-informed embedding space that supervises the audio encoder, motivating a mechanism that identifies and filters such chunks at both training and inference.
Our approach builds on a key insight of~\citet{Bara_ski_2025}: ASR hallucinations are not arbitrary but structurally predictable, drawn from a small and stable vocabulary of training-data artifacts. This predictability lets us turn hallucination detection into a learnable task, implemented as a binary classifier (Figure~\ref{fig:overview_livi}.a).  Its design is grounded in the hypothesis that the information making a chunk hallucination-prone is already encoded in the ASR encoder's latent features~\citep{glazer2025transcriptionmechanisticinterpretabilityasr}, since the decoder attends to these when generating hallucinations.

\paragraph{Shared ASR backbone (feature extractor)}
We first define a shared ASR backbone that is used as a frozen feature extractor, providing latent representations $z_i$ that are used as inputs by all branches of the framework described next.
Given a 30-second audio excerpt $x_i$ from a track $x \in \mathcal{C}$, we first extract these latent features via the frozen ASR encoder $f_{\text{extract}}$:
$$
f_{\text{extract}} :
\begin{cases}
\mathcal{C} \;\to\; \mathbb{R}^{L \times d_w} \\
x_i \;\mapsto\; z_i
\end{cases}
$$

\paragraph{Lyrics-Informed Embedding Space (teacher)}
We now define the lyrics-informed embedding space, which serves as supervision for training the audio encoder.
Given latent features $z_i$ extracted via the ASR encoder $f_{\text{extract}}$, we first complete transcription through the ASR decoder $f_{\text{transc}}$. The corresponding lyrics embedding $t_i \in \mathbb{R}^d$ is then obtained by encoding this transcript with a pre-trained text embedding model $f_{\text{text}}$. This two-step process can be expressed as a fixed encoder:
$$
g_{\text{text}} = f_{\text{text}} \circ f_{\text{transc}} :
\begin{cases}
\mathbb{R}^{L\times d_w} \;\to\; \mathbb{R}^d \\
z_i \;\mapsto\; t_i
\end{cases}
$$
\paragraph{Audio Encoder (student)}
Next, we define an audio encoder that projects latent states $z_i$ into the lyrics-informed embedding space via a projection head $f_{\text{proj}}$, yielding the audio embedding $a_i \in \mathbb{R}^d$:
$$
g_{\text{audio}} = f_{\text{proj}} :
\begin{cases}
\mathbb{R}^{L\times d_w} \;\to\; \mathbb{R}^d \\
z_i \;\mapsto\; a_i
\end{cases}
$$

\paragraph{Binary classifier} 
Finally, we introduce a binary classifier $h_{\text{halluc}}$ that outputs the probability $p_{\text{halluc}}$ that a chunk is hallucination-prone. The chunk is discarded when $p_{\text{halluc}}$ is greater than a threshold $\delta$:
$$
h_{\text{halluc}} :
\begin{cases}
\mathbb{R}^{L\times d_w} \;\to\; [0, 1] \\
z_i \;\mapsto\; p_{\text{halluc}}
\end{cases}
$$

\newpage
\paragraph{Alignment Loss}
The main objective is to learn $g_{\text{audio}}$ such that audio embeddings $a_i$ are aligned with their corresponding lyric embeddings $t_i$ under cosine similarity. Formally, this corresponds to minimizing the loss:
$$
\mathcal{L}_{\text{cos}} = \sum_{z_i \in B} \Big(1 - s\big(g_{\text{audio}}(z_i), g_{\text{text}}(z_i)\big)\Big)
$$
 
Yet the training objective can be pushed further given the data available: since the lyrics-informed space is fixed and the target lyrics embeddings are accessible during training, one can leverage not only pointwise audio-text alignment but also the geometry of the target space. More specifically, the inter-sample distances between lyrics embeddings can serve as an additional supervision signal to guide the training of the audio encoder. 
In principle, enforcing pointwise alignment alone is sufficient to reproduce the target space, provided enough training data and iterations: if every audio embedding is correctly aligned with its corresponding text embedding, the inter-sample geometry follows as a consequence. In practice, however, explicitly enforcing this geometry within each training batch provides a more direct and efficient supervision signal~\citep{Park_2019_CVPR}. This also acts as a regularizer, guarding against representation collapse by preserving inter-sample separation.
Formally, we enforce that pairwise similarities between audio embeddings, $s(a_i, a_j)$, match those of the corresponding lyrics embeddings, $s(t_i, t_j)$.
Given a batch $\{(z_i, t_i)\}_{i=1}^{B}$, this component of the training objective takes the form:
\[
\mathcal{L}_{\text{MSE}}
= \frac{1}{B^{2}} \sum_{i,j=1}^{B}
\Bigl(s(a_i, a_j) - s(t_i, t_j)\Bigr)^{2}.
\]
This yields the final objective optimized during training, combining a pointwise alignment term $\mathcal{L}_{\text{cos}}$ with a geometry-preservation term $\mathcal{L}_{\text{MSE}}$:
\[
\mathcal{L}_{\text{align}}
= \alpha \,\mathcal{L}_{\text{cos}}
+ (1-\alpha)\,\mathcal{L}_{\text{MSE}},
\qquad \alpha \in [0,1]
\]

\paragraph{Classifier Loss}
We train the binary classifier as a standard classification task using binary cross-entropy against the weak labels $y_i \in \{0,1\}$:
\[
\mathcal{L}_{\text{BCE}}
= -\frac{1}{B}\sum_{i=1}^{B}
\Big[\, y_i \log p_{\text{halluc}}(z_i) + (1-y_i)\log\big(1 - p_{\text{halluc}}(z_i)\big)\Big]
\]



\subsection{Two-Stage Retrieval at Inference}
\label{sec:inference}

At inference, retrieval proceeds in two stages and relies on both local and global representations (Figure~\ref{fig:inference_pipeline}): global embeddings first generate a candidate set, which is then reranked by the MaxSim metric over local embeddings, borrowing from the late-interaction literature~\citep{khattab2020colbert}.


A query track $q$ is first divided into $N_q$ fixed-length, partially overlapping chunks $\{x_i\}_{i=1}^{N_q}$. Each chunk is encoded by the ASR backbone into latent features $z_i = f_{\text{extract}}(x_i)$ and kept only if $h_{\text{halluc}}(z_i) < \delta$. Each retained chunk is then projected by $g_{\text{audio}}$ into a local embedding $a_i$, forming the set
\[
A_q = \{a_i  \mid h_{\text{halluc}}(z_i) < \delta\}.
\]
Their $\ell_2$-normalized average yields the global embedding
\[
\mathbf{e}_q = \frac{\bar{a}_q}{\lVert \bar{a}_q \rVert_2}, \qquad \bar{a}_q = \frac{1}{|A_q|}\sum_{i=1}^{|A_q|} a_i
\]
\begin{figure}[t!]
\centering
\resizebox{\linewidth}{!}{
\begin{tikzpicture}[
  font=\small,
  >={Latex[length=2.2mm]},
  node distance=8mm,
  box/.style={draw, rounded corners, align=center, minimum height=13mm, text width=21mm, inner sep=3pt, fill=gray!6},
  db/.style={draw, cylinder, shape border rotate=90, aspect=0.25, minimum height=12mm, minimum width=17mm, align=center, fill=gray!6, font=\footnotesize},
  lbl/.style={font=\footnotesize, align=center}
]
\node[box] (track) {Query\\track $q$};
\node[box, right=of track, text width=24mm] (extract) {Segment into 30s \& encode $f_{\text{extract}}$};
\node[box, right=of extract] (filt) {Filter\\$h_{\text{halluc}} < \delta$};
\node[box, above=9mm of filt, text width=16mm, minimum height=9mm] (stop) {Discard};
\node[box, right=of filt] (proj) {Project\\$g_{\text{audio}}$};
\node[box, right=15mm of proj] (s1) {\textbf{Stage 1}\\ball search\\on $\mathbf{e}_q$};
\node[box, right=of s1] (s2) {\textbf{Stage 2}\\MaxSim rerank\\on $A_q$};
\node[box, right=of s2] (out) {Ranked\\cover list};
\node[db, below=11mm of s1] (cat) {Catalog\\embeddings};

\draw[->] (track) -- (extract);
\draw[->] (extract) -- node[lbl, above]{$z_i$} (filt);
\draw[->] (filt) -- (stop);
\draw[->] (filt) -- node[lbl, above]{$z_i'$} (proj);
\draw[->] (proj) -- node[lbl, above]{$A_q,\ \mathbf{e}_q$} (s1);
\draw[->] (s1) -- node[lbl, below]{$B_\tau(q)$} (s2);
\draw[->] (s2) -- (out);
\draw[->] (cat) -- (s1);

\begin{scope}[on background layer]
\node[draw, dashed, rounded corners, inner sep=7pt, fit=(s1)(s2),
      label={[font=\footnotesize\itshape]above:Two-Stage Retrieval}] {};
\end{scope}
\end{tikzpicture}
}
\caption{\textbf{Inference pipeline.} A query track $q$ is segmented into overlapping $30\,$s chunks ($10\,$s overlap) and encoded by the ASR backbone $f_{\text{extract}}$ into latent features $z_i$. Chunks with $h_{\text{halluc}}(z_i) \geq \delta$ are discarded; the remaining chunks are projected by $g_{\text{audio}}$ into a set of local embeddings $A_q$, whose $\ell_2$-normalized average yields the global embedding $\mathbf{e}_q$. Stage~1 performs a ball search over the catalog's global embeddings, returning a candidate set $B_\tau(q)$. Stage~2 reranks these candidates with the MaxSim late-interaction score $S_{\max}(q,x)$ over the local embeddings, producing the final ranking.}
\label{fig:inference_pipeline}
\end{figure}

\paragraph{Stage 1 --- Candidate generation}A ball search over global embeddings retrieves the catalog tracks whose global similarity to the query exceeds a threshold $\tau$:
\[
B_\tau(q) \;=\; \bigl\{\, x \in \mathcal{C}\setminus\{q\} \; \mid \; \cos(\mathbf{e}_q, \mathbf{e}_x) \ge \tau \,\bigr\}.
\]

We prefer ball search over $k$-nearest-neighbour for two reasons. First, cover classes are expected to vary widely in size, so a fixed number of neighbours would either miss versions from large classes or over-retrieve for small ones, whereas the threshold $\tau$ should adapt the candidate set to each query (see Section~\ref{sec:reranking}). Second, ball search better reflects our assumption that the semantic lyric similarity between versions is strictly bounded.

\paragraph{Stage 2 --- Late-interaction reranking.} 
Each candidate $x \in B_\tau(q)$ is rescored by a MaxSim late-interaction operator over local embeddings~\citep{khattab2020colbert}, defined as:

\[
S_{\max}(q,x) \;=\; \frac{1}{|A_q|}\sum_{a \in A_q}\; \max_{b \in A_x}\; \cos(a,b).
\]

For every query chunk embedding $a_i \in A_q$, we take its similarity to the best-matching candidate chunk embedding in $A_x$. These maxima are then averaged over all query chunks to obtain the final score $S_{\max}(q,x)$. This operator confers robustness to structural reordering, partial overlap, and differing track lengths---variations characteristic of covers that averaging into a single global vector would dilute (Sections~\ref{sec:reranking} and~\ref{sec:maxsim_interp}).




\section{Implementation Details}
\label{sec:livi}

\subsection{Lyrics-Informed Embedding Space}
\label{sec:space}
The lyrics-informed embedding space (Figure~\ref{fig:overview_livi}.b) is defined as 
\[\mathcal{T} = \{ g_{\text{text}}(z_i)  \mid z_i = f_{\text{extract}}(x_i),\ x_i \in \mathcal{C} \},
\]
where $g_{\text{text}} = f_{\text{text}} \circ f_{\text{transc}}$ maps latent features $z_i$ to a lyric embedding $t_i \in \mathbb{R}^d$ via transcription. This space clusters semantically similar lyrics---assumed to represent versions---closer together than unrelated ones, and serves as the target structure for training the audio encoder $g_{\text{audio}}$.

\paragraph{Transcription model: Encoder $f_{\text{extract}}$ and Decoder $f_{\text{transc}}$}
We adopt Whisper as our ASR model for three reasons that align with our objectives:
(i)~trained on data spanning 96 languages, it offers broad multilingual coverage;
(ii)~it demonstrates strong robustness to noise, accents, and real-world acoustic
variability~\citep{radford2022robustspeechrecognitionlargescale, Gong_2023}, making it well-suited to the heterogeneous conditions of music audio;
and (iii)~its open-source release facilitates seamless integration into our pipeline. We use \textit{whisper-large-v3-turbo}\footnote{\url{https://huggingface.co/openai/whisper-large-v3-turbo}},
an optimized variant of \textit{whisper-large-v3} that offers comparable transcription accuracy with up to $8\times$ faster inference. Its architecture combines a convolutional front-end, 32 Transformer encoder layers, and a 4-layer Transformer decoder that generates transcriptions autoregressively~\citep{radford2022robustspeechrecognitionlargescale}.

\paragraph{Text embedding model $f_{\text{text}}$} 
As mentioned in Section~\ref{sec:methodo}, the objective is to construct an embedding space for lyric transcriptions that is efficient for our downstream task, providing a structure for the audio encoder to align with. 
Recent progress in Natural Language Processing has produced text embedding models especially well suited to this purpose. Most are derived from large pre-trained language models and fine-tuned for semantic similarity, typically within the Sentence-BERT framework~\citep{reimers2019sentencebertsentenceembeddingsusing}, which maps full sentences into fixed-size embeddings that preserve semantic proximity. 

Based on an evaluation of six multilingual text embedding models for the downstream task (see Section~\ref{sec:val-space}), we select \textit{gte-multilingual-base}\footnote{\url{https://huggingface.co/Alibaba-NLP/gte-multilingual-base}}~\citep{zhang2024mgtegeneralizedlongcontexttext}, an encoder-only Transformer that produces 768-dimensional embeddings across more than 70 languages and achieves SOTA results in multilingual retrieval for models of comparable size~\citep{zhang2024mgtegeneralizedlongcontexttext}. As shown in Section~\ref{sec:val-space}, it outperforms alternatives in our evaluations, with performance approaching the ceiling defined by editorial lyrics, underscoring both its robustness to transcription noise and its suitability as the backbone of our pipeline. In our case, we rely on this off-the-shelf model without additional fine-tuning, as it already provides strong results. We leave task-specific fine-tuning for future work, where it could further enhance performance.

\subsection{Shared ASR backbone}
\label{sec:backbone}
We adopt the encoder of the Whisper ASR model as our audio backbone $f_{\text{extract}}$ for both the audio encoder and the binary classifier. Given a 80-channel log-Mel spectrogram, the encoder produces a sequence of hidden states $z\in \mathbb{R}^{L \times d_w}$, where $d_w = 1280$ is the latent dimension and $L=1500$ the number of frames for a 30s input.

For the audio encoder, this choice is motivated by two considerations: (i) Whisper internal representations, shaped by the ASR training objective, are expected to capture phonetic and linguistic information~\citep{glazer2025transcriptionmechanisticinterpretabilityasr} that makes them suitable for alignment with lyrics embeddings; and (ii) reusing the encoder from the same ASR model used to build the target embedding space enables our model to specifically target the decoder in order to have an efficient alternative. Consequently, we keep the encoder frozen to preserve this alignment and maintain the latent structure learned during Whisper’s large-scale training. For the binary classifier, since Whisper hallucinations arise from the decoder attending to the encoder's hidden states, we hypothesize that the information that makes a chunk hallucination-prone should already be present in those representations~\citep{glazer2025transcriptionmechanisticinterpretabilityasr}. This choice also serves efficiency, a key objective throughout our design: reusing the frozen ASR encoder keeps the classifier lightweight, adding negligible inference cost beyond the forward pass already required for the audio encoder.

\subsection{Audio Encoder}
\label{sec:audio_encoder}

\begin{figure}[!h]
\centering
\includegraphics[width=0.7\linewidth]{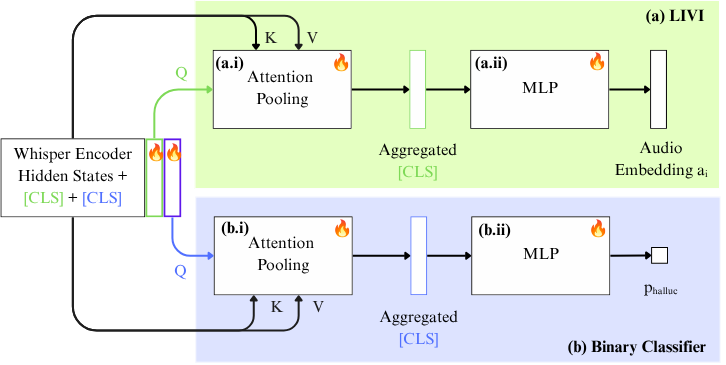}
\caption{\textbf{Architecture of the audio encoder $g_{\text{audio}}$ and the binary classifier $h_{\text{halluc}}$.} Given Whisper encoder hidden states, two separate [CLS] tokens are appended and processed through two parallel branches. \\ 
{ \small
\textbf{(a) LIVI:} \textbf{(i)} An attention pooling module aggregates frame-level representations using the first [CLS] token as query. \textbf{(ii)} A multi-layer perceptron then projects the pooled representation into the lyrics-informed embedding space, yielding the final audio embedding $a_i$. \\ 
\textbf{(b) Binary Classifier:} \textbf{(i)} A second attention pooling module aggregates frame-level representations using the second [CLS] token as query. \textbf{(ii)} A classification head then estimates the hallucination probability $p_{\text{halluc}} $.
}}
\label{fig:model_architecture}
\end{figure}

The LIVI audio encoder $g_{\text{audio}}$ (Figure~\ref{fig:overview_livi}.c) is defined as
\[
g_{\text{audio}} = f_{\text{proj}} : z_i \mapsto a_i, \qquad z_i = f_{\text{extract}}(x_i),\ x_i \in \mathcal{C},
\]
mapping latent features $z_i$ to an embedding $a_i \in \mathbb{R}^d$ aligned with its lyric-based counterpart $t_i = g_{\text{text}}(z_i)$. 
Taking Whisper encoder hidden states $z_i \in \mathbb{R}^{L\times d_w}$ as input, an attention-based pooling module aggregates these frame-level representations into a single vector, which a projection head then maps into the lyrics-informed embedding space. By refining Whisper's latent space rather than learning cross-modal alignment from scratch, this design keeps the model compact and computationally efficient during both training and inference.
Figure~\ref{fig:model_architecture}.a details this architecture.

\paragraph{Attention-based temporal pooling (Fig.~\ref{fig:model_architecture}.a.i)}
To reduce frame-level representations into a fixed-dimensional vector suitable for projection into the lyrics-informed embedding space, we adopt an attention-based pooling mechanism inspired by~\citet{touvron2021patchconvnet}. A learnable
$[\text{CLS}]$ token $q_{\text{cls}} \in \mathbb{R}^{d_w}$ is appended to the hidden states $H$ and acts as the sole query in a single-head attention mechanism with Rotary Positional Embeddings (RoPE)~\citep{su2023roformerenhancedtransformerrotary}.
Formally, given the projections
\[
Q = q_{\text{cls}} W_Q, \qquad K = H W_K, \qquad V = H W_V,
\]
where $W_Q, W_K, W_V \in \mathbb{R}^{d_w \times d_k}$ are learnable matrices and $d_k$ denotes the dimensionality of the key vector (set to $d_w$ in our implementation), the pooled representation is
\[
h = \mathrm{Attention}(Q, K, V)
  = \mathrm{softmax}\!\left(\frac{QK^\top}{\sqrt{d_k}}\right) V \;\in \mathbb{R}^{d_w}.
\]

The attention weights determine the relative importance of each frame, guiding how information from the sequence is aggregated into the [CLS] representation. This representation is then passed through a residual feed-forward block with LayerNorm to yield the final pooled embedding $h \in \mathbb{R}^{d_w}$.

\paragraph{Projection Head (Fig.~\ref{fig:model_architecture}.a.ii)}
The pooled vector $h$ is finally projected into the lyrics-informed embedding space ($d=768$)
through a four-layer MLP with hidden sizes [3072, 2048, 2048, 1536], a configuration selected based on empirical validation, totaling 13.6M trainable parameters. Each intermediate layer is followed by LayerNorm and a ReLU activation, while the final layer outputs the audio embedding $a_i$ used in the downstream retrieval task.

\subsection{Binary classifier}
\label{sec:denoiser}

\subsubsection{Architecture} 

The binary classifier $h_{\text{halluc}}: z_i \;\mapsto\; p_{\text{halluc}}$ (Figure~\ref{fig:overview_livi}.a) estimates the probability that a chunk yields transcription hallucinations, used to filter such chunks out of the pipeline.
Like the audio encoder (Section~\ref{sec:audio_encoder}), the classifier operates directly on the shared latent features $z_i$ produced by Whisper's frozen encoder. An attention-pooling module (Figure~\ref{fig:model_architecture}.b.i), trained independently from the audio encoder, aggregates these frame-level hidden states into a fixed-size representation, which is then passed to a classification head (Figure~\ref{fig:model_architecture}.b.ii), implemented as an MLP with hidden sizes [512, 256, 128]. Each intermediate layer is followed by LayerNorm and a ReLU activation, while the final layer outputs a $2$-dimensional vector over which a softmax yields a per-chunk hallucination probability $p_{\text{halluc}}$.

\subsubsection{Weak-Label Creation}
\label{sec:weak_labels} 
Training the binary classifier requires per-chunk labels, which we derive through a two-step procedure. Our approach builds on a key insight of~\citet{Bara_ski_2025}: ASR hallucinations are not arbitrary but structurally predictable, drawn from a small and stable vocabulary of training-data artifacts.
Building on this, we first identify the hallucinations Whisper produces most frequently on non-vocal music audio and compile them into a reference set $\mathcal{H}$. Second, each chunk is labeled hallucination-prone if its transcription, once stripped of phrases from this set, remains too short, and reliable otherwise. Both steps operate on chunks flagged as instrumental by a vocal-detection model, which we describe first.

\paragraph{Vocal-detection model} We rely on a proprietary vocal-detection model to flag 30-second audio chunks as instrumental.
Built on the Musicnn backbone~\citep{pons2019musicnnpretrainedconvolutionalneural}\footnote{\scriptsize Model architecture and pretrained weights are available at \url{https://github.com/jordipons/musicnn}.}, augmented with a binary classification head, it estimates a vocalness probability for each 3-second audio window, and a global chunk-level score is then obtained by averaging across windows.  
Although this step relies on a proprietary vocal-detection model, which limits strict reproducibility, the model is used only offline to construct $\mathcal{H}$, and this dependency does not affect the final pipeline. As an unexplored alternative, one could consider exploiting an open vocal annotation source such as DALI~\citep{https://doi.org/10.5281/zenodo.1492443}.



\paragraph{Step 1: Building a Bag of Hallucinations}
\label{sec:step1_boh}
We first identify the hallucinations Whisper produces most frequently on instrumental music.
Chunks flagged as instrumental by the vocal-detection model are transcribed with Whisper, normalized, and the resulting phrases counted across the corpus (see Section~\ref{sec:setup} for dataset details).
Following~\citet{Bara_ski_2025}, we form the Bag of Hallucinations $\mathcal{H}$ by retaining phrases exceeding a frequency threshold $k$. 
Since some frequent hallucinations coincide with phrases that could plausibly appear in genuine lyrics (e.g \textit{``i love you''} or \textit{``and''}), we apply a manual pass to remove such entries, ensuring $\mathcal{H}$ captures spurious artifacts rather than legitimate lyrical content.
We set the frequency threshold to $k=10$. At this threshold, $\mathcal{H}$ contains 429 phrases, covering 95.3\% of hallucination occurrences in the instrumental corpus.
The most common hallucinations, with their occurrences, are reported in Appendix~\ref{app:boh}. Notably, \textit{``thank you''} alone accounts for over half of all hallucination occurrences (55.68\%) in our corpus (1,023,917 chunks).

\begin{figure}[h!]
    \centering
    \includegraphics[width=0.45\textwidth]{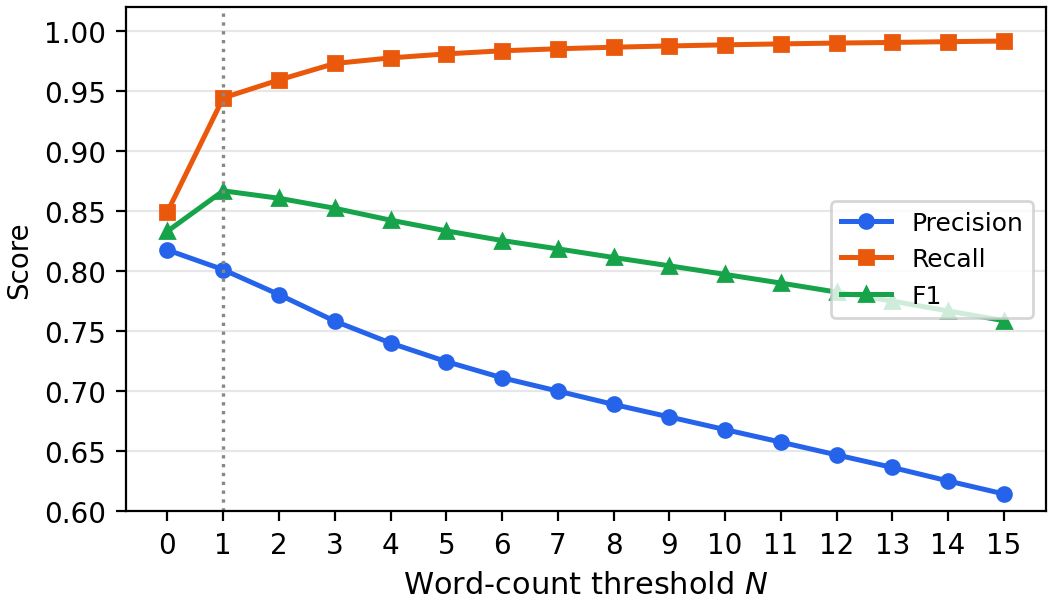}
    \caption{\textbf{Sensitivity of chunk-level weak labels to the word-count threshold $N$.} Precision, recall, and F1 are computed by comparing the reliable/hallucination-prone labels---defined by whether the length of the cleaned transcription is below $N$---against the proprietary vocalness model's labels, on an instrumental set.}
\label{fig:N_sensitivity}
\end{figure}

\paragraph{Step 2: Deriving Chunk-level Labels}
\label{sec:step2_boh}
Next, we assign each training chunk a reliable or hallucination-prone tag: its transcription is normalized as in Step~1 and stripped of any phrase present in $\mathcal{H}$, then flagged hallucination-prone if the result contains fewer than $N$ words, and reliable otherwise. The threshold $N$ is set by evaluating the resulting reliable/hallucination-prone labels against the proprietary vocalness model's outputs on the instrumental set. Figure~\ref{fig:N_sensitivity} reports precision, recall, and F1 as $N$ varies: recall drops sharply at $N=0$, while precision degrades gradually as $N$ grows beyond 1. We select $N=1$, which achieves the best F1 (0.867, with precision 0.801 and recall 0.944).

\section{Experimental Setup}
\label{sec:setup}
 
\subsection{Training Data}
\label{sec:training_data}

We use a subset $\mathcal{D}$ of $643,342$ tracks from Discogs-VI~\citep{araz2024discogsvimusicalversionidentification}, linked to a proprietary catalog to recover the corresponding .mp3 audio. Each track is segmented into audio chunks $x_i$ of 30 seconds with 10-second overlap, then log-Mel spectrograms $\omega_i$ are precomputed using Whisper's feature extractor. For the binary classifier, labels $y_i$ are derived as described in Section~\ref{sec:weak_labels}, and training is restricted to a balanced subset of $1.5$M pairs $(z_i, y_i)$. For the LIVI audio encoder, hallucination-prone chunks are discarded and lyric embeddings $t_i = g_{\mathrm{text}}(x_i) \in \mathbb{R}^d$ are derived through the frozen text pipeline of Section~\ref{sec:space}. Pairs are randomly restricted to $1.5$M entries to reduce training time. Both subsets follow an 80/10/10 train/validation/test split at the clique level, preventing leakage across partitions, resulting in around $1.2$M training pairs, $150$k eval/test pairs.

\subsection{Training Configuration}
\label{sec:training_config}

Both models are trained on a single NVIDIA RTX A5000 (24\,GB) for $3$ epochs with batch size $128$ ($\approx33$ hours), using AdamW (weight decay = $0.01$, $\beta=(0.9, 0.98)$) with a fixed learning rate of $10^{-4}$ and linear warmup over the first $10$k steps for the LIVI audio encoder and $2$k steps for the binary classifier. Mixed precision is enabled for the Whisper encoder, while the rest of the model is trained in full precision. Early stopping is applied based on the average cosine similarity between audio and text embeddings on the validation set for the audio encoder, and on F1-score for the binary classifier. 

\subsection{Benchmarks}
\label{sec:benchmarks}
 
We evaluate on the standard Covers80~\citep{ellis2007covers80} and  SHS100k-TEST~\citep{bertin2011million} benchmarks, together with  the test set of Discogs-VI~\citep{araz2024discogsvimusicalversionidentification}, restricted to entries with an  available YouTube link. All datasets follow the same preprocessing  pipeline: tracks are first retrieved from YouTube using the provided  links and matched against a proprietary catalog via audio fingerprinting;  tracks for which \emph{all} chunks are flagged as hallucination-prone  by the binary classifier (Section~\ref{sec:denoiser}) are then discarded. Catalog coverage varies across datasets, and is notably lower for SHS100k, whose matched subset retains only 2,727 of the original 7,108 tracks (38\%). Results on this benchmark should therefore be interpreted with some caution regarding their generalization to the full SHS100k.
We restrict evaluation queries to tracks belonging to cliques with at least two versions, ensuring each query has a true match. Retrieval is performed against the full dataset, so singleton-clique tracks remain in the catalog as noise. 
Table~\ref{tab:dataset_stats} reports, for each dataset, the initial number of tracks and cliques, along with the counts after catalog matching, vocal filtering, and query selection. It also reports the resulting clique size distribution.

At this point, it should be acknowledged that our lyrics-centered method excludes some tracks from the reference datasets.
Consequently, $100$\% of Covers80, $61.33$\% of SHS100k and $97.85$\% of Discogs-VI are retained for evaluation. 
We leave as future work the use of different musical modalities to retrieve these instances.
Nevertheless, our experiments operate at a large scale of  $80{,}485$ tracks for Discogs-VI, yielding an evaluation setting that reflects real-world conditions.

\begin{table}[h!]
    \centering
    \small
    \resizebox{\textwidth}{!}{%
    \begin{tabular}{l rr rrr rrrrrr}
        \toprule
        & \multicolumn{2}{c}{\textbf{Initial}} 
        & \multicolumn{4}{c}{\textbf{After matching}}
        & \multicolumn{5}{c}{\textbf{Clique size}} \\
        \cmidrule(lr){2-3}\cmidrule(lr){4-7}\cmidrule(lr){8-12}
        \textbf{Dataset} 
        & \textbf{Tracks} & \textbf{Cliques}
        & \textbf{Tracks} & \textbf{Vocal Tracks} & \textbf{Queries} & \textbf{Cliques}
        & \textbf{Avg} & \textbf{Std} & \textbf{Med} & \textbf{Min} & \textbf{Max} \\
        \midrule
        Covers80   & 160 & 80 & 123 & 123    & 92     & 46     & 2.0   & 0.0   & 2.0  & 2.0 & 2.0   \\
        SHS100K    & 7,108 & 111 & 2,727 & 1,672  & 1,671  & 104    & 23.2  & 30.5  & 12.0 & 2.0 & 220.0 \\
        Discogs-VI & 98,785 & 493,049 & 82,247 & 80,485 & 53,162 & 16,508 & 3.22  & 2.81  & 2.0  & 2.0 & 63.0  \\
        \bottomrule
    \end{tabular}%
    }
\caption{\textbf{Test dataset statistics before and after preprocessing.} \textit{Initial} reports the raw number of tracks and cliques in each dataset. \textit{After matching} reports counts following matching with our proprietary catalog: \textit{Tracks} = tracks successfully matched to the catalog; \textit{Vocal Tracks} = matched tracks with detected vocal content; \textit{Queries} = vocal tracks belonging to cliques with $\geq 2$ versions; \textit{Cliques} = number of such cliques, with size statistics (Avg, Std, Med, Min, Max) reported over the same set.}
\label{tab:dataset_stats}
\end{table}
 
\subsection{Evaluation Configuration}
\label{sec:metrics}
 
We adopt a standard retrieval protocol for version identification~\citep{yesiler2020accuratescalableversionidentification}. Following the retrieve-then-rerank pipeline introduced in Section~\ref{sec:inference}, candidates are first retrieved via cosine similarity on global embeddings, with the ball search radius set to $\tau = 0.85$ (the choice of $\tau$ is discussed in Section~\ref{sec:reranking}). They are then reranked using the MaxSim score on local embeddings. Retrieval performance is assessed using standard metrics: MR1, the mean rank of the first true positive; HR@1, the fraction of queries with the correct cover ranked first; and MAP@10, which evaluates precision within the top 10 results.

\section{Experiments}
\label{sec:results}
 
\subsection{Validation of the Lyrics-Informed Embedding Space}
\label{sec:val-space}

We evaluate the lyrics embeddings $t_i=g_{\text{text}}(z_i)$ in the downstream retrieval task to assess the performance of the lyrics-informed embedding space. 
Results are reported in Table~\ref{tab:gte_clean_vs_transc} across six multilingual text embedding models $f_{\text{text}}$: \textit{gte-multilingual-base}~\citep{zhang2024mgtegeneralizedlongcontexttext}, \textit{multilingual-e5-\{small, large, large-instruct\}}~\citep{wang2024multilinguale5textembeddings}, \textit{jina-embeddings-v3}~\citep{sturua2024jinaembeddingsv3multilingualembeddingstask} and \textit{multilingual-mpnet-base-v2}~\citep{reimers2019sentencebertsentenceembeddingsusing}.
We assess the models both on editorial lyrics embeddings  $\tilde{t}_i = f_{\text{text}}(\ell_i)$ and on transcribed lyrics embeddings $t_i = f_{\text{text}} (f_{\text{transc}}(z_i))$, thus evaluating the models both with and without transcription noise. 
Consequently, this evaluation is limited to subsets of the datasets for which editorial lyrics are available in our proprietary catalog, yielding 116, 438, and 4,623 tracks for Covers80, SHS100k, and Discogs-VI, respectively.

\begin{table}[b!]
\scriptsize
\centering
\begin{tabular}{l ccc ccc ccc}
    \toprule
    & \multicolumn{3}{c}{\textbf{Covers80}} 
    & \multicolumn{3}{c}{\textbf{SHS100k}} 
    & \multicolumn{3}{c}{\textbf{Discogs-VI}} \\
    \cmidrule(lr){2-4}\cmidrule(lr){5-7}\cmidrule(lr){8-10}
    \textbf{Model} 
    & MR1$\downarrow$ & HR1$\uparrow$ & MAP$\uparrow$
    & MR1$\downarrow$ & HR1$\uparrow$ & MAP$\uparrow$
    & MR1$\downarrow$ & HR1$\uparrow$ & MAP$\uparrow$ \\
    \midrule
    \multicolumn{10}{l}{\textit{(i) Editorial lyrics}} \\
    \quad gte-b      & \textbf{1.000} & \textbf{1.000} & \textbf{1.000} & \textbf{2.393} & \textbf{0.956} & \textbf{0.922} & \textbf{12.16} & \textbf{0.934} & \textbf{0.913} \\
    \quad e5-s       & \textbf{1.000} & \textbf{1.000} & \textbf{1.000} & 3.779 & 0.947 & 0.902 & 21.22 & 0.924 & 0.895 \\
    \quad e5-l       & \textbf{1.000} & \textbf{1.000} & \textbf{1.000} & 6.032 & 0.954 & 0.916 & 19.53 & 0.919 & 0.897 \\
    \quad e5-l-inst  & \textbf{1.000} & \textbf{1.000} & \textbf{1.000} & 3.223 & 0.949 & 0.908 & 16.96 & 0.923 & 0.901 \\
    \quad jina       & \textbf{1.000} & \textbf{1.000} & \textbf{1.000} & 5.211 & 0.951 & 0.921 & 17.44 & 0.924 & 0.905 \\
    \quad mpnet      & 4.139 & 0.924 & 0.921 & 4.772 & 0.947 & 0.892 & 39.96 & 0.902 & 0.861 \\
    \midrule
    \multicolumn{10}{l}{\textit{(ii) Transcribed lyrics}} \\
    \quad gte-b      & 1.101 & 0.975 & 0.987 & \textbf{2.046} & \textbf{0.971} & \textbf{0.942} & \textbf{13.21} & \textbf{0.929} & 0.893 \\
    \quad e5-s       & 1.025 & 0.975 & 0.987 & 3.954 & 0.954 & 0.908 & 14.56 & 0.927 & 0.890 \\
    \quad e5-l       & 1.051 & 0.975 & 0.983 & 3.299 & 0.954 & 0.913 & 15.33 & 0.922 & 0.891 \\
    \quad e5-l-inst  & \textbf{1.013} & \textbf{0.987} & \textbf{0.994} & 2.233 & 0.951 & 0.914 & 13.44 & 0.926 & \textbf{0.897} \\
    \quad jina       & 1.089 & 0.975 & 0.977 & 2.485 & 0.966 & 0.935 & 13.81 & 0.925 & 0.886 \\
    \quad mpnet      & 3.367 & 0.899 & 0.914 & 2.199 & 0.954 & 0.908 & 40.95 & 0.878 & 0.800 \\
    \bottomrule
\end{tabular}
\caption{\textbf{Music Cover Retrieval results across six text embedding models.}
We report results using both editorial lyrics embeddings and transcribed lyrics embeddings. 
For each column, bold indicates the best result within each category.
}
\label{tab:gte_clean_vs_transc}
\end{table}

Among the six candidates, \textit{gte-multilingual-base} emerges as the most effective since it ranks first across nearly all metrics on Discogs-VI, the largest and most representative benchmark of real-world conditions. Its performance on transcribed lyrics approaches the ceiling defined by editorial lyrics, underscoring both its robustness to transcription noise and its suitability as the backbone of our pipeline.
Additionally, results show that most multilingual encoders achieve competitive scores and, notably, reach ceiling performance on Covers80 when provided with editorial lyrics. This underscores the role of lyrics as a stable and discriminative signal for Music Cover Retrieval and validates the lyrics-informed embedding space as a robust supervisory signal: it captures the semantic structure necessary to distinguish versions. In our case, we rely on off-the-shelf multilingual models without additional fine-tuning, as they already provide strong results. We leave task-specific fine-tuning for future work, where it could further enhance performance.

\subsection{Alignment of Audio with Lyrics-Informed Embeddings}
\label{sec:audio_alignment}

We now examine whether the audio encoder $g_{\text{audio}}$ successfully maps audio inputs into the lyrics-informed embedding space. Concretely, we assess the alignment between audio embeddings $a_i = g_{\text{audio}}(z_i)$ and their paired text targets $t_i = g_{\text{text}}(z_i)$ on the held-out test split (Section~\ref{sec:training_data}), at both chunk and track levels. Track-level embeddings, denoted $\bar{a}_u$ and $\bar{t}_u$, are obtained by averaging chunk-level embeddings from a track $u$, followed by $\ell_2$-normalization.
This yields $153{,}884$ chunk-level and $19{,}670$ track-level pairs.
Table~\ref{tab:cosim_audio_text} reports distribution metrics for the pointwise cosine similarities $\cos(a_i,t_i)$ between paired audio and text embeddings. 
At chunk level, the model achieves a mean similarity of $0.903$ ($\sigma = 0.037$), which rises to $0.957$ ($\sigma = 0.028$) at track level. While the higher mean and lower variance at the track level suggest that the encoder integrates local cues into stable global representations, the results further demonstrate that our approach achieves tight audio-lyric alignment.

\begin{figure}[!h]
    \centering
    \begin{minipage}{0.8\linewidth}
        \centering
        \begin{subfigure}[t]{0.48\linewidth}
            \centering
            \includegraphics[width=\linewidth]{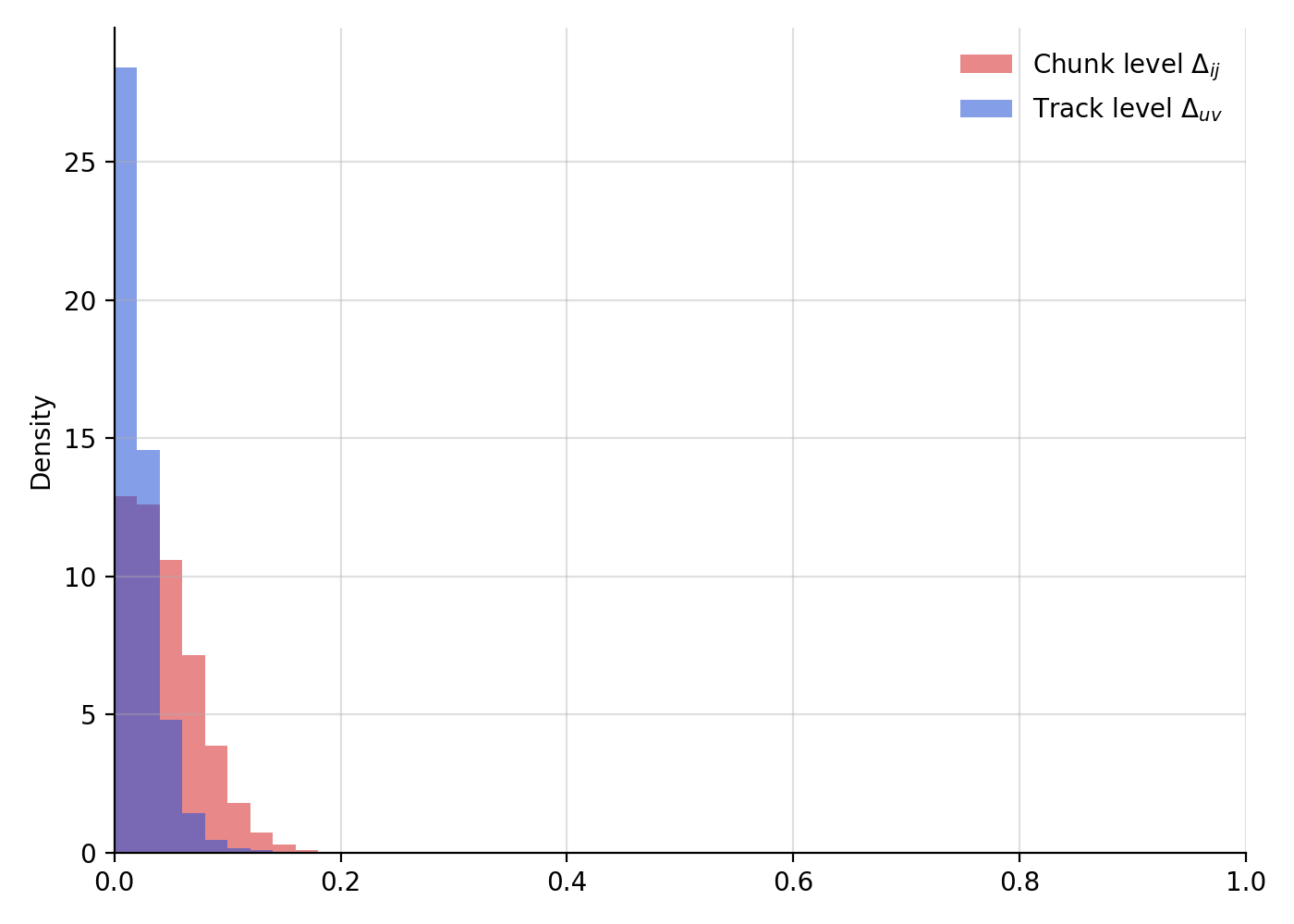}
            \caption{Distributions of pairwise distance differences $\Delta_{ij}$
            and $\Delta_{uv}$.}
            \label{fig:pairwise_audio}
        \end{subfigure}
        \hfill
        \begin{subfigure}[t]{0.48\linewidth}
            \centering
            \includegraphics[width=\linewidth]{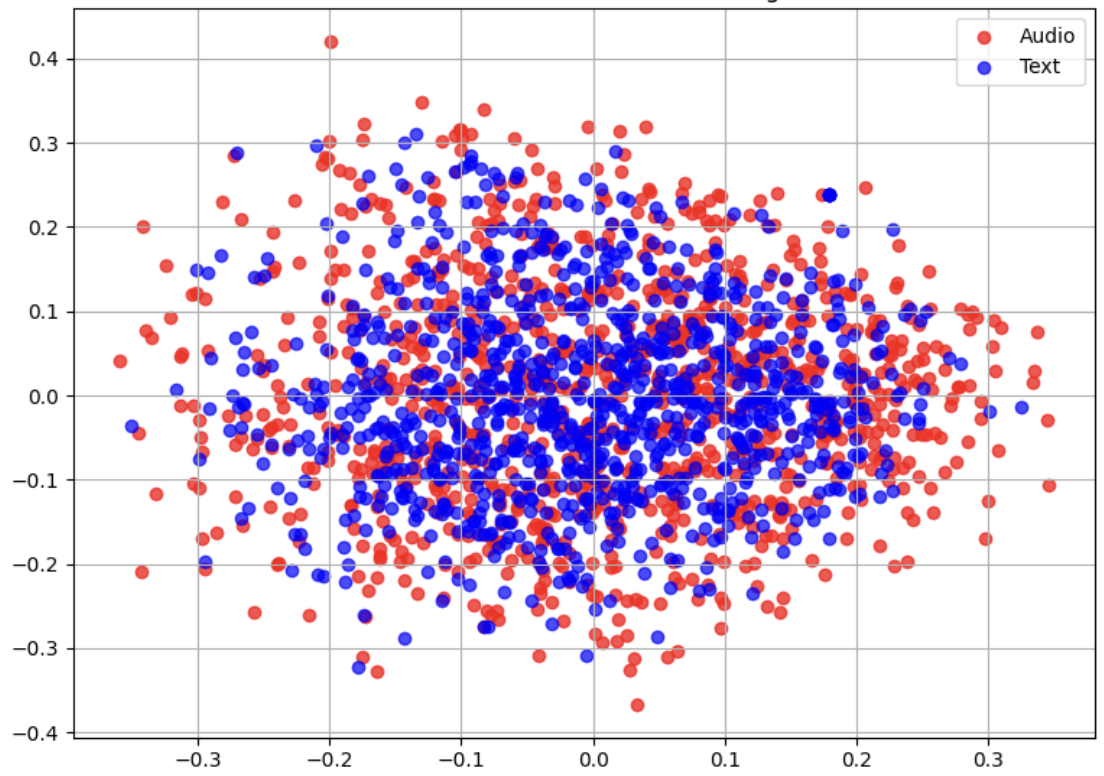}
            \caption{PCA projection of audio and text
            embeddings for $1{,}000$ random tracks.}
            \label{fig:pca_audio}
        \end{subfigure}
    \end{minipage}
    \caption{Geometric alignment between audio embeddings 
    and text embeddings.}
    \label{fig:audio_alignment}
\end{figure}

\begin{table}[h!]
    \centering
    \small
    \begin{tabular}{lccccc}
        \toprule
        & \textbf{Mean} & \textbf{Std} & \textbf{Median} 
        & \textbf{Min} & \textbf{Max} \\
        \midrule
        Chunk-level & $0.903$ & $0.037$ & $0.909$ & $0.557$ & $0.979$ \\
        Track-level & $0.957$ & $0.028$ & $0.965$ & $0.632$ & $0.991$ \\
        \bottomrule
    \end{tabular}
    \caption{Pointwise cosine similarity between audio embeddings 
    $a_i = g_{\text{audio}}(z_i)$ and paired text embeddings 
    $t_i = g_{\text{text}}(z_i)$, at chunk and track levels.}
    \label{tab:cosim_audio_text}
\end{table}
Figure~\ref{fig:pairwise_audio} complements this with the distribution of pairwise distance differences  $\Delta_{ij} = |\cos(a_i, a_j) -  \cos(t_i, t_j)|$ at the chunk level and $\Delta_{uv} = |\cos(\bar{a}_u, \bar{a}_v) - \cos(\bar{t}_u, \bar{t}_v)|$ at the track level, providing a global picture of how well the geometric structure of the target textual embedding space is preserved. 
The distributions are sharply concentrated near zero, indicating that the audio encoder not only aligns individual embeddings with their targets but also preserves the relational structure of the textual space.
The PCA projection of $1{,}000$ randomly sampled tracks (Figure~\ref{fig:pca_audio}) further confirms cross-modal alignment, with both modalities spanning across overlapping regions in the shared embedding space.

\subsection{Validation of the Data Processing Pipeline}
\label{sec:denoiser-eval}

\begin{table}[h!]
\small
\centering
\begin{subtable}{0.48\textwidth}
\centering
\begin{tabular}{l ccc}
    \toprule
    & \textbf{Precision} & \textbf{Recall} & \textbf{F1} \\
    \midrule
    Reliable            & 0.9775 & 0.8187 & 0.8911 \\
    Halluc.-prone        & 0.8469 & 0.9815 & 0.9093 \\
    \midrule
    Macro avg.           & 0.9122 & 0.9001 & 0.9002 \\
    Weighted avg.        & 0.9115 & 0.9011 & 0.9004 \\
    \bottomrule
\end{tabular}
\caption{Per-class classification results.}
\label{tab:denoiser_clf}
\end{subtable}%
\hfill
\begin{subtable}{0.48\textwidth}
\centering
\begin{tabular}{l cc}
    \toprule
    & \multicolumn{2}{c}{\textbf{Predicted}} \\
    \cmidrule(lr){2-3}
    \textbf{Actual} & Reliable & Halluc.-prone \\
    \midrule
    Reliable       & 61,014 & 13,508 \\
    Halluc.-prone  & 1,404  & 74,768 \\
    \bottomrule
\end{tabular}
\caption{Confusion matrix.}
\label{tab:denoiser_confmat}
\end{subtable}
\caption{Per-class classification results of the binary classifier on the held-out test set (150,694 samples), at the decision threshold $p_{\text{halluc}} = 0.5$.}
\label{tab:denoiser_eval}
\end{table}

\begin{figure}[b!]
    \centering
    \begin{minipage}{0.8\linewidth}
        \centering
        \begin{subfigure}[b]{0.48\linewidth}
            \includegraphics[width=\linewidth]{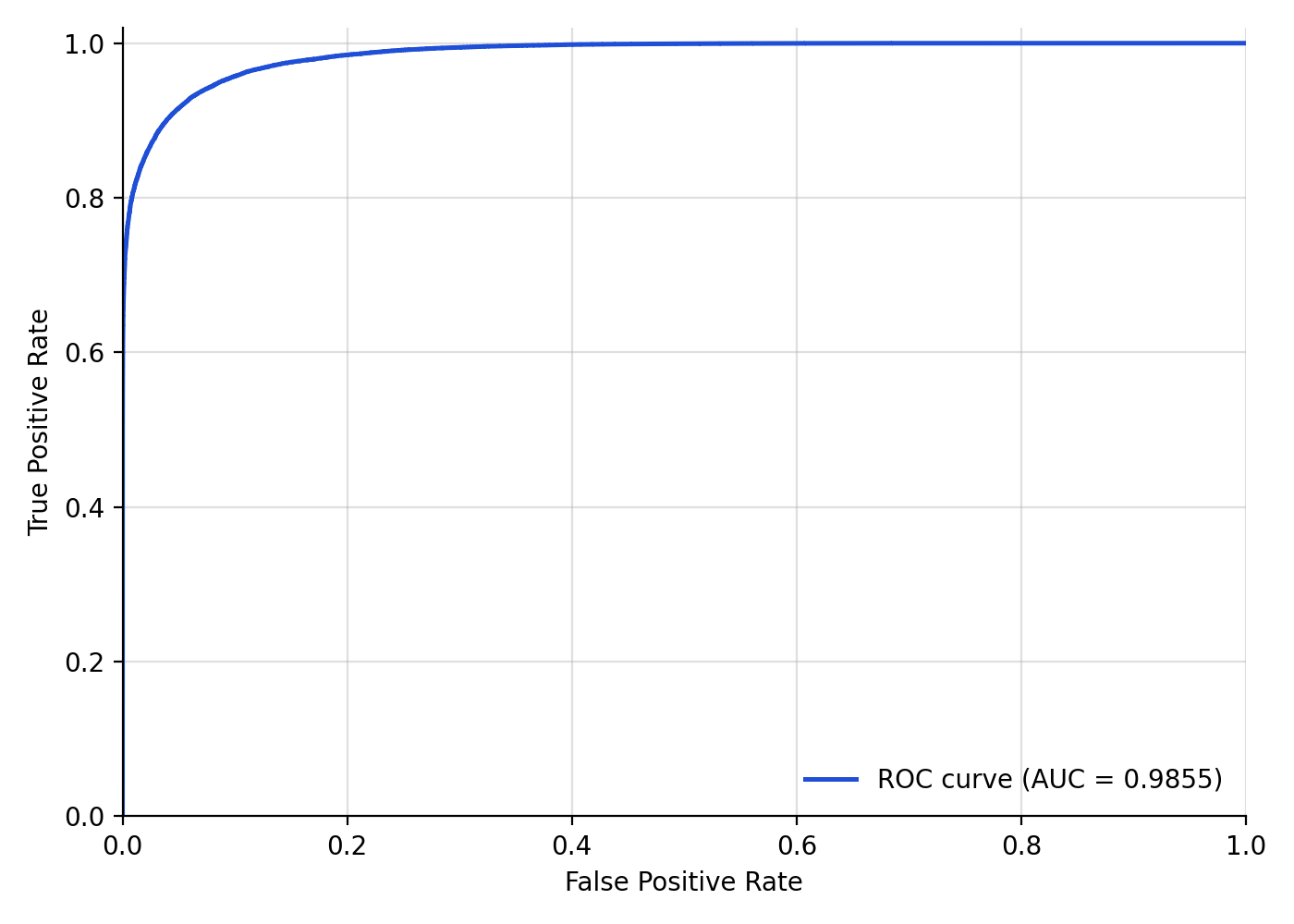}
                \caption{ROC curve.}

            \label{fig:denoiser_roc}
        \end{subfigure}
        \hfill
        \begin{subfigure}[b]{0.48\linewidth}
            \includegraphics[width=\linewidth]{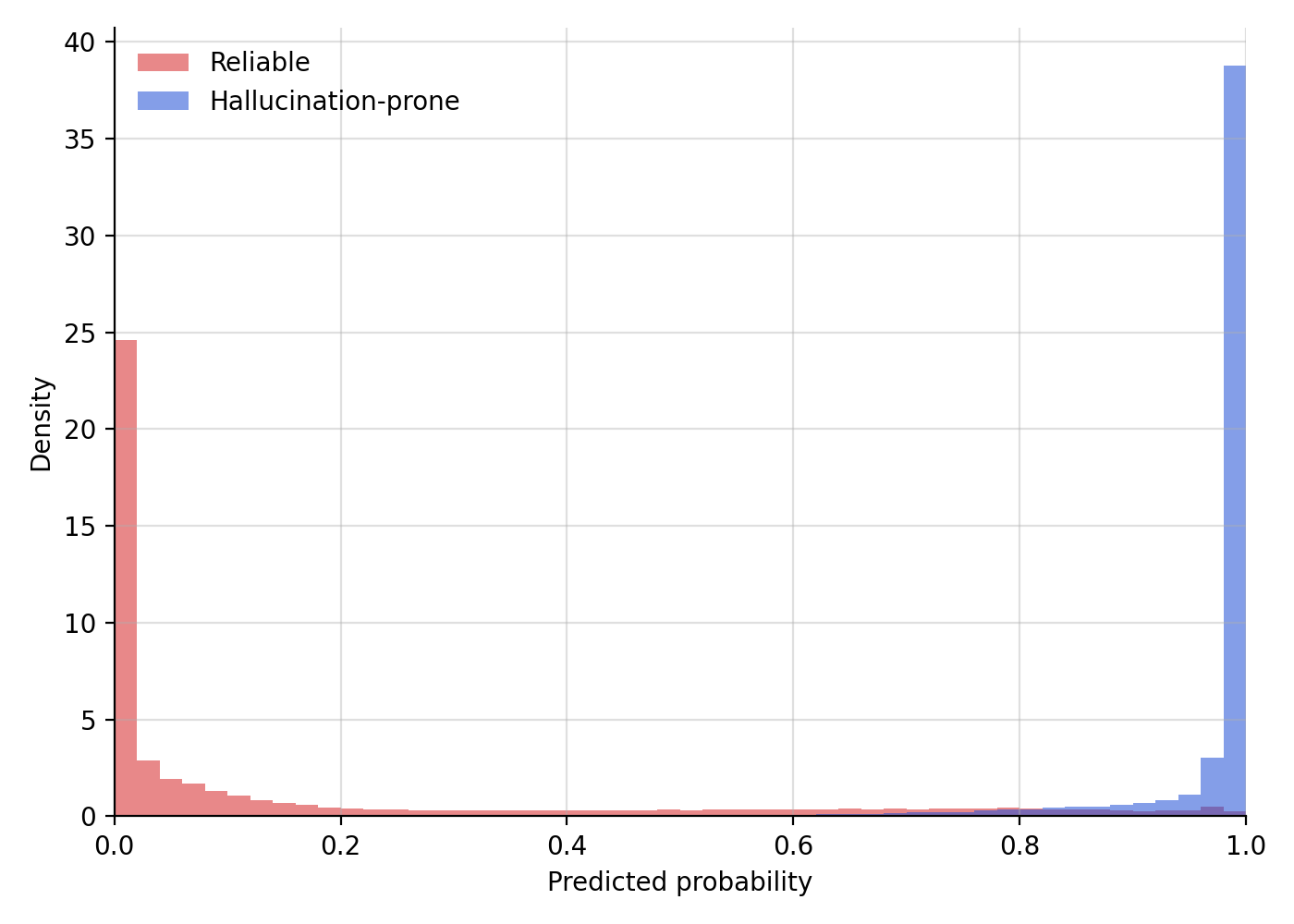}
            \caption{Predicted probability distribution across classes.}
            \label{fig:denoiser_prob}
        \end{subfigure}
    \end{minipage}
    \caption{\textbf{Binary classifier evaluation on the held-out test set.} The decision threshold is set at $p_{\text{halluc}} = 0.5$.}
    \label{fig:denoiser_eval}
\end{figure}

We evaluate the binary classifier from Section~\ref{sec:denoiser} as a standard classification task. Table~\ref{tab:denoiser_clf} reports per-class precision, recall and F1-score on the held-out test set (Section~\ref{sec:setup}), along with macro and weighted averages across both classes. ROC curve and predicted probability distributions across classes are shown in Figure~\ref{fig:denoiser_eval}.
At the chosen threshold, the classifier achieves high recall on the Hallucination-prone class (0.9815), catching nearly all true hallucinations, at the cost of a lower precision (0.8469). We consider this trade-off acceptable, as discarding a borderline chunk is preferable to admitting a hallucinated one. 
The ROC curve (Figure~\ref{fig:denoiser_roc}) achieves an AUC of 0.9855, indicating strong separability between classes. The predicted probability distributions (Figure~\ref{fig:denoiser_prob}) are sharply concentrated near 0 and 1 with little overlap at the decision boundary, confirming a confident, well-separated classifier.

Next, we assess whether filtering out hallucination-prone chunks yields more reliable transcriptions and, in turn, more faithful target embeddings. Two complementary metrics are reported in Table~\ref{tab:denoiser_wer}: the Word Error Rate (WER) against editorial lyrics, measuring how accurately transcriptions capture the actual lyric content, and the number of BoH phrases (Section~\ref{sec:weak_labels}) detected in the transcriptions, quantifying the residual presence of known hallucination artifacts. Both are reported on the full test set and on the subset retained by the binary classifier (chunks with $p_{\text{halluc}} < 0.5$), restricted to chunks with available lyrics in our proprietary catalog. This yields 32,427
chunks for the full test set and 18,755 chunks for the classifier-filtered subset.
An effective filter should reduce both quantities, indicating fewer hallucinations propagating through the text pipeline.
WER against editorial lyrics drops from 0.6570 to 0.4630 (-30\%), and BoH occurrences fall from 157 to 22 (-86\%). 
These results confirm that filtering out hallucination-prone chunks yields more reliable transcriptions, and that the classifier reliably isolates chunks responsible for known hallucination artifacts.

\begin{table}[t!]
\centering
\small
\begin{tabular}{l cccc}
    \toprule
    \textbf{Subset} & \textbf{\# chunks} & \textbf{WER} $\downarrow$ & \textbf{BoH count} $\downarrow$ \\
    \midrule
    Full test set                      & 32,427 & 0.6570 & 157 \\
    Retained by binary classifier      & 18,755 & 0.4630 & 22  \\
    \bottomrule
\end{tabular}
\caption{WER against editorial lyrics and Bag of Hallucinations (BoH) occurrence counts across subsets of the
binary classifier test set, restricted to the subset of chunks with editorial lyrics found in our proprietary catalog.}
\label{tab:denoiser_wer}
\end{table}

\subsection{Validation of the Two-Stage Retrieval Procedure}
\label{sec:reranking}

\paragraph{First Stage --- Ball Search vs.\ $k$-NN}

\begin{table}[b!]
\centering
\small
\begin{tabular}{l r rrrr}
    \toprule
    \textbf{Method} & \textbf{Recall} $\uparrow$ &
    \textbf{Mean} & \textbf{Std} & \textbf{Median} & \textbf{Max} \\
    \midrule
    $k$-NN ($k=100$)        & 0.9160 & 100  & ---  & 100  & 100  \\
    $k$-NN ($k=1000$)       & 0.9491 & 1000 & ---  & 1000 & 1000 \\
    \midrule
    Ball search $\tau=0.83$ & 0.9489 & 1940.56 & 2601.84 & 852  & 24,711 \\
    Ball search $\tau=0.85$ & 0.9255 & 688.16 & 1139.48 & 199 & 13,069 \\
    Ball search $\tau=0.88$ & 0.8852 & 86.19  & 213.22  & 14   & 3,221  \\
    Ball search $\tau=0.90$ & 0.8501 & 14.59  & 46.55   & 3    & 1,366  \\
    Ball search $\tau=0.95$ & 0.6174 & 1.76   & 3.71    & 0    & 48    \\
    \bottomrule
\end{tabular}
\caption{First-stage retrieval comparison: recall and distribution of per-query candidate set size (mean, standard deviation, median, and max) on the Discogs-VI training set (209,049 tracks).}
\label{tab:ball_search}
\end{table}

The first stage defines the candidate set $B_\tau(q)$ for reranking. Its recall therefore upper-bounds the full pipeline, while its size governs reranking cost.
We compare ball search with $\tau \in \{0.83, 0.85, 0.88, 0.90, 0.95\}$ against k-NN with $k \in \{100, 1000\}$ on the Discogs-VI training set (209,049 tracks, see Section~\ref{sec:setup}), reporting recall and per-query candidate set size (Table~\ref{tab:ball_search}).
We argue that ball search with $\tau = 0.85$ yields the best recall/size trade-off, achieving a recall of $0.9255$ with a median of $199$ candidates per query ($\mu = 688.16$, $\sigma = 1139.48$), close to the recall of $k$-NN ($k=1000$) at a fraction of the candidate-set size.
It is also better suited to our setting: candidate set size follows from the global embedding geometry rather than a fixed a priori value, adapting to a realistic catalog where version clique sizes are likely heterogeneous, ranging from pairs to hundreds of recordings (see Table~\ref{tab:dataset_stats} for clique size distribution across benchmarks).

\paragraph{Second Stage --- MaxSim Reranking}

To characterize the behaviour of MaxSim reranking, we track, for every positive pair $(q, v^+)$ with $v^+ \in B_{0.85}(q)$, how the rank of the true cover $v^+$ shifts from cosine similarity to MaxSim scoring, comparing $\mathrm{rank}_{s(q,v^+)}$ against $\mathrm{rank}_{S_{\max}(q,v^+)}$.
For each benchmark, we partition all positive pairs returned by ball search into four bins according to their stage-1 rank. Table~\ref{tab:maxsim_bucket} reports, for each bin, the fraction of pairs improved or degraded by MaxSim, the magnitude of these changes, and the fraction of degraded pairs for which another member of the same clique is still ranked above them.

\begin{table}[!b]
\centering
\small
\resizebox{\columnwidth}{!}{%
\begin{tabular}{ll r ccc ccc}
\toprule
& & & \multicolumn{3}{c}{\textbf{Improved}} & \multicolumn{3}{c}{\textbf{Degraded}} \\
\cmidrule(lr){4-6} \cmidrule(lr){7-9}
\textbf{Benchmark} & \textbf{Stage-1 rank} & \textbf{n} & \textbf{\%} & \textbf{med.\ gain} & \textbf{IQR} & \textbf{\%} & \textbf{med.\ drop} & \textbf{\% higher} \\
\midrule
\multirow{2}{*}{Covers80}
  & rank $=1$    & 89 & \phantom{0}0.0 & --- & --- & \phantom{0}0.0 & --- & --- \\
  & rank 2--10   & 3  & 100.0          & 2   & 1.5--2.5 & \phantom{0}0.0 & --- & --- \\
\midrule
\multirow{4}{*}{SHS100k}
  & rank $=1$    & 1{,}587  & \phantom{0}0.0 & --- & --- & 56.5 & \phantom{0}3 & 99.8 \\
  & rank 2--10   & 12{,}179 & 29.6           & 2  & 1--4   & 55.3 & \phantom{0}5 & 100.0 \\
  & rank 11--100 & 41{,}566 & 45.0           & 12 & 5--24  & 50.5 & 15 & 100.0 \\
  & rank $>100$  & 21{,}530 & 58.5           & 26 & 8--53  & 38.4 & 14 & 100.0 \\
\midrule
\multirow{4}{*}{Discogs-VI}
  & rank $=1$    & 44,261 & \phantom{0}0.0 & ---   & ---         & 22.8 & \phantom{0}1 & 95.1 \\
  & rank 2--10   & 113,591 & 34.6           & \phantom{0}2 & 1--3   & 32.9 & \phantom{0}2 & 98.7 \\
  & rank 11--100 & 43,090  & 56.0           & \phantom{0}5 & 2--11  & 30.7 & \phantom{0}3 & 98.3 \\
  & rank $>100$  & \phantom{0} 6,074 & 77.7 & 216          & 112--522 & 22.2 & 186 & 83.3 \\
\bottomrule
\end{tabular}}
\caption{\textbf{Impact of MaxSim reranking per stage-1 rank bin, across benchmarks.}\\
\emph{Improved}: fraction of pairs, median and inter-quartile range (IQR) rank gain for pairs improved under MaxSim.\\
\emph{Degraded}: fraction of pairs, median rank drop and percentage of degraded pairs that have another cover ranked better in their clique.}
\label{tab:maxsim_bucket}
\end{table}

MaxSim benefits low-ranked covers most. On Discogs-VI, for instance, among pairs ranked 2-10 by stage-1 cosine similarity, $34.6\%$ see their rank improved, and this goes up up to $77.7\%$ for stage-1 ranks beyond 100. We can see a similar effect for SHS100K, with $58.5\%$ of stage-1 ranks beyond 100 improving under MaxSim.
Gain magnitude follows the same trend, with median improvements rising from +2 positions in the top 10 to +216 and +26 positions beyond rank 100 for Discogs-VI and SHS100k. This supports our hypothesis that chunk-level matching is most valuable where global embeddings lack the resolution to discriminate between versions: low-ranked covers sit in a dense region where cosine similarity alone cannot disambiguate, yielding a large candidate set on which MaxSim is most effective. 
Where degradation does occur, it is largely harmless: in 83.3–100.0\% of degraded pairs, another clique member outranks them after MaxSim. This means the query is still likely to retrieve a correct cover at high rank, with MaxSim simply promoting a different version of the same track.

\subsection{Downstream Evaluation on Music Cover Retrieval}
\label{sec:main_results}

We evaluate the proposed LIVI pipeline on the music cover retrieval task (Table~\ref{tab:vi_main}), following the protocol of Section~\ref{sec:metrics}. Results are organised into three groups:
\paragraph{(i) System variants} We introduce a baseline obtained by mean-pooling frame-level Whisper encoder states; this removes the projection module and training stage, thereby isolating the effect of LIVI’s alignment beyond raw ASR features.
To isolate the contribution of MaxSim reranking (Section~\ref{sec:inference}), we also report LIVI (no MaxSim): the output of the first retrieval stage alone, ranking candidates by cosine similarity on track-level embeddings without the second-stage MaxSim rerank.

\paragraph{(ii) Lyrics embeddings} 
Since the LIVI audio encoder is trained to align with the lyrics embedding space, these textual representations can serve as approximate upper bounds for what audio-based retrieval can achieve. To get a sense of the headroom remaining in our pipeline, we report three variants: $t_{\text{local}}$, which averages segment-level text embeddings; $t_{\text{local+MaxSim}}$, which applies MaxSim reranking over these, thus mimicking LIVI's full retrieval process from the textual embedding space; and $t_{\text{global}}$, where tracks are not segmented and a single embedding is computed directly from the full-track transcription.
\paragraph{(iii) Audio baselines} We compare against recent competitive methods: \textit{ByteCover2}~\citep{bytecover2}, \textit{CLEWS}~\citep{serrà2025supervisedcontrastivelearningweaklylabeled}, \textit{CQTNet}~\citep{9053839}, and \textit{DViNet} ~\citep{araz2024discogsvinet_mirex}, evaluated using official implementations and pretrained checkpoints from~\citet{serrà2025supervisedcontrastivelearningweaklylabeled}. As with the baselines, whose embeddings are computed at the local level, we derive global embeddings in the same manner as for LIVI and evaluate all methods under the protocol of Section~\ref{sec:metrics} for a fair comparison. We report results using MaxSim reranking on one hand, and global embeddings alone on the other hand (no MaxSim).

\begin{table}[!h]
\small
\centering
\resizebox{\textwidth}{!}{%
\begin{tabular}{l ccc ccc ccc}
    \toprule
    & \multicolumn{3}{c}{\textbf{Covers80}} 
    & \multicolumn{3}{c}{\textbf{SHS100k}} 
    & \multicolumn{3}{c}{\textbf{Discogs-VI}} \\
    \cmidrule(lr){2-4}\cmidrule(lr){5-7}\cmidrule(lr){8-10}
    \textbf{Method} 
    & MR1$\downarrow$ & HR1$\uparrow$ & MAP$\uparrow$
    & MR1$\downarrow$ & HR1$\uparrow$ & MAP$\uparrow$
    & MR1$\downarrow$ & HR1$\uparrow$ & MAP$\uparrow$ \\
    \midrule
    \multicolumn{10}{l}{\textit{(i) System variants}} \\
    \quad LIVI                        & \textbf{1.00} & \textbf{1.000} & \textbf{1.000} & 5.31 & 0.961 & 0.906 & 642.35 & \underline{0.845} & \underline{0.811} \\
    \quad LIVI (no MaxSim)          & \underline{1.07} & 0.967 & 0.979 & 6.67$^{\dagger}$ & 0.957 & 0.902$^{\dagger}$ & 644.45$^{\dagger}$ & 0.833$^{\dagger}$ & 0.794$^{\dagger}$ \\
    \quad Mean-pooling                     & 13.03$^{\dagger}$ & 0.565$^{\dagger}$ & 0.635$^{\dagger}$ & 9.89$^{\dagger}$ & 0.906$^{\dagger}$ & 0.722$^{\dagger}$ & 1560.93$^{\dagger}$ & 0.537$^{\dagger}$ & 0.394$^{\dagger}$ \\
    \midrule
    \multicolumn{10}{l}{\textit{(ii) Lyrics embeddings}} \\
    \quad $t_{\text{local+MaxSim}}$   & \underline{1.07} & \underline{0.978} & \underline{0.983} & 5.46 & 0.967 & 0.925$^{\dagger}$ & \textbf{356.81}$^{\dagger}$ & \textbf{0.847}$^{\dagger}$ & \textbf{0.819}$^{\dagger}$ \\
    \quad $t_{\text{local}}$          & 1.86 & 0.934 & 0.942 & 8.41$^{\dagger}$ & 0.961 & 0.908 & 480.47$^{\dagger}$ & 0.824$^{\dagger}$ & 0.789$^{\dagger}$ \\
    \quad $t_{\text{global}}$         & 1.20 & 0.945 & 0.964 & 5.46 & 0.967 & 0.926$^{\dagger}$ & \underline{356.92}$^{\dagger}$ & 0.837$^{\dagger}$ & 0.806$^{\dagger}$ \\
    \midrule
    \multicolumn{10}{l}{\textit{(iii) Audio baselines}} \\
    \quad ByteCover2 & 2.33 & 0.913 & 0.921 & \underline{1.35}$^{\dagger}$ & \underline{0.989}$^{\dagger}$ & 0.968$^{\dagger}$ & 413.32$^{\dagger}$ & 0.791$^{\dagger}$ & 0.739$^{\dagger}$ \\
    \quad ByteCover2 (no MaxSim)& 1.89 & 0.902 & 0.918 & \textbf{1.30}$^{\dagger}$ & \underline{0.989}$^{\dagger}$ & 0.967$^{\dagger}$ & 413.32$^{\dagger}$ & 0.791$^{\dagger}$ & 0.738$^{\dagger}$\\

    \quad CLEWS     & 1.60 & 0.945 & 0.963 & 1.54$^{\dagger}$ & \textbf{0.991}$^{\dagger}$ & \textbf{0.980}$^{\dagger}$ & 612.81$^{\dagger}$ & 0.811$^{\dagger}$ & 0.784$^{\dagger}$ \\
    \quad CLEWS (no MaxSim)    & 2.16 & 0.869$^{\dagger}$ & 0.900 & 1.69$^{\dagger}$ & 0.988$^{\dagger}$ & \underline{0.973}$^{\dagger}$ & 614.40$^{\dagger}$ & 0.793$^{\dagger}$ & 0.755$^{\dagger}$ \\

    \quad CQTNet   & 4.43$^{\dagger}$ & 0.793$^{\dagger}$ & 0.829$^{\dagger}$ & 3.20 & 0.947 & 0.873$^{\dagger}$ & 1345.13$^{\dagger}$ & 0.580$^{\dagger}$ & 0.485$^{\dagger}$ \\
    \quad CQTNet (no MaxSim)      & 3.98$^{\dagger}$ & 0.793$^{\dagger}$ & 0.827$^{\dagger}$ & 3.07 & 0.952 & 0.879$^{\dagger}$ & 1344.62$^{\dagger}$ & 0.582$^{\dagger}$ &  0.487$^{\dagger}$\\

    \quad DViNet    & 3.17 & 0.869$^{\dagger}$ & 0.891 & 2.33$^{\dagger}$ & 0.975 & 0.950$^{\dagger}$ & 811.79$^{\dagger}$ & 0.691$^{\dagger}$ & 0.635$^{\dagger}$ \\
    \quad DViNet (no MaxSim)     & 2.85 & 0.880$^{\dagger}$ & 0.902 & 2.28$^{\dagger}$ & 0.976 & 0.952$^{\dagger}$ & 811.52$^{\dagger}$ & 0.692$^{\dagger}$ & 0.636$^{\dagger}$ \\

    \bottomrule
\end{tabular}%
}
\caption{\textbf{Music Cover Retrieval results.} \\ 
{\small 
\textbf{(i)} \textbf{LIVI:} proposed full pipeline (Section~\ref{sec:inference}); \textbf{LIVI (no MaxSim):} first retrieval stage only, ranking by global embedding similarity without MaxSim reranking; \textbf{Mean-pooling:} mean-pooled encoder states. \\
\textbf{(ii)} \textbf{Lyrics embeddings:} $t_{\text{local+MaxSim}}$ (MaxSim reranking), $t_{\text{local}}$ (without MaxSim reranking), $t_{\text{global}}$ (full transcription). \\
\textbf{(iii)} 
\textbf{Audio baselines}~\citep{serrà2025supervisedcontrastivelearningweaklylabeled, bytecover2, 9053839, araz2024discogsvinet_mirex}. Bold: best; underline: second-best. $^{\dagger}$: statistically significant difference to LIVI.}}
\label{tab:vi_main}
\end{table}

As shown in Table~\ref{tab:vi_main}, LIVI matches or exceeds the performance of lyrics embeddings across all metrics and datasets, except for MR1 on Discogs-VI. 
Mean-pooled embeddings, by contrast, yield poor performance across all datasets, underscoring the gains from LIVI’s architecture and training strategy. Through attention-based pooling and a projection network, Whisper’s outputs are effectively mapped into the lyrics-informed embedding space, producing representations sufficiently discriminative to recognize versions. 
Across all three benchmarks, MaxSim reranking improves retrieval for both the LIVI and lyrics embeddings, relative to ranking by global cosine similarity alone. This is consistent with our motivation for late interaction: the lyrics-informed embeddings are trained at the chunk level, so local matching exploits a level of granularity the global average discards. 

Compared to audio baselines, LIVI delivers competitive or superior performance in our benchmark, with particularly strong results on Covers80 and Discogs-VI, where it outperforms all in HR@1 and MAP@10 (Table~\ref{tab:vi_main}). LIVI does lag behind on MR1 for Discogs-VI; however, MR1 can be disproportionately skewed by a few severely mis-ranked queries, and such tail effects warrant cautious interpretation.
On SHS100k, the audio baselines lead, with CLEWS attaining the highest scores (HR@1 $= 0.991$, MAP@10 $= 0.980$) and ByteCover2 following closely ($0.989$, $0.968$). This difference can be partially explained by dataset characteristics: SHS100k includes a notable fraction of challenging cases for our lyrics-centered approach (see Section~\ref{sec:breakdown}). 
While differences on Covers80 are often not statistically significant, likely due to limited dataset size and ceiling effects across models, LIVI achieves statistically significant improvements over all audio baselines across two out of three metrics on the largest and most diverse benchmark Discogs-VI ($p < 0.05$, Holm-Bonferroni-corrected; Wilcoxon Signed-Rank for MR1/MAP, McNemar's test for HR@1).

Together, these results show that LIVI generalizes effectively to Music Cover Retrieval. It matches the lyrics-based upper bound, clearly outperforms raw mean-pooled Whisper representations, and competes with or surpasses state-of-the-art audio baselines on vocal tracks. Its relatively simple and reproducible design contrasts with the complexity of models like ByteCover2, yet its performance across datasets establishes LIVI as a compact and powerful alternative.

\subsection{Efficiency: Model Size and Inference Time}
\label{sec:efficiency}

\begin{figure}[!t]
\centering
\includegraphics[width=0.5\linewidth]{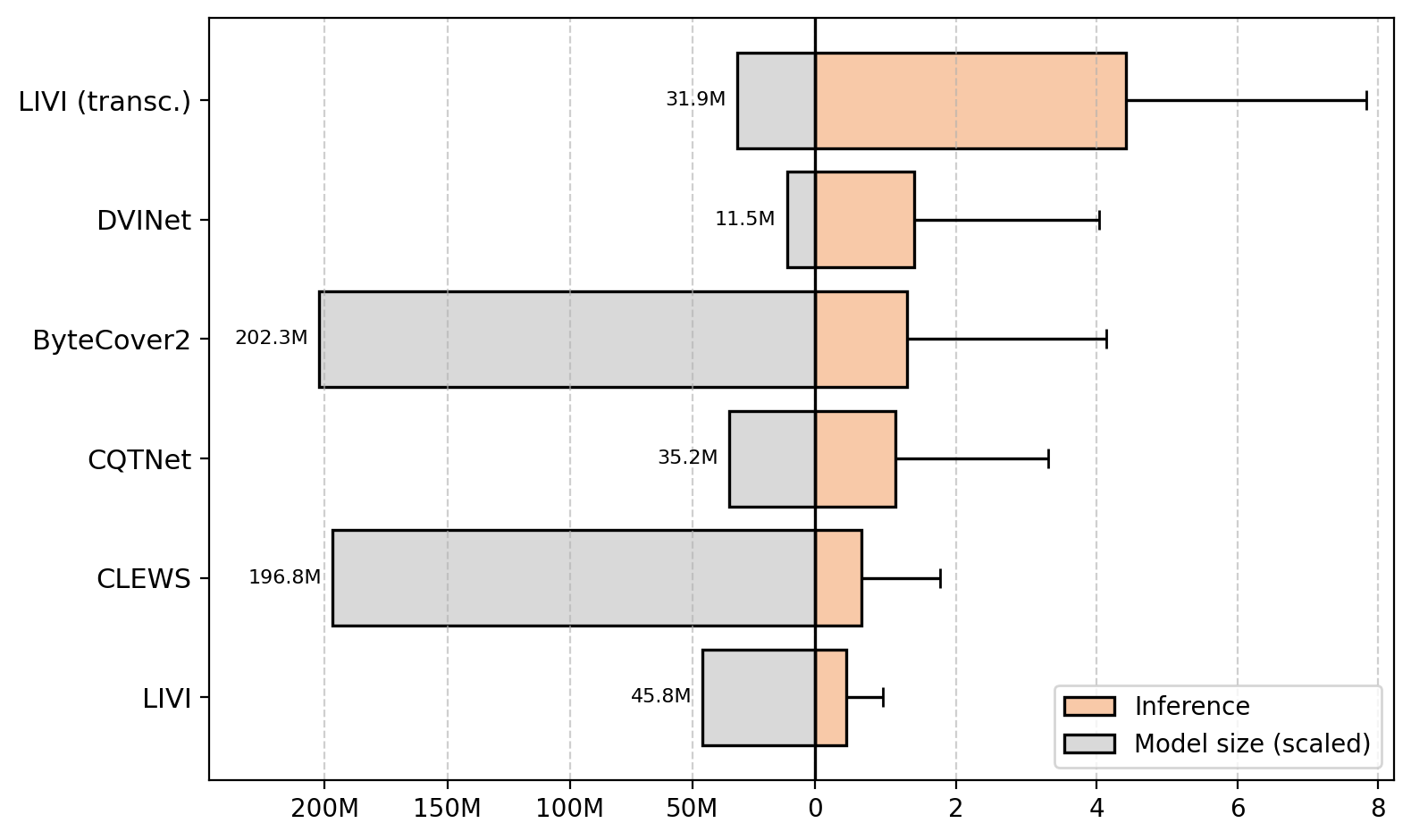}
\caption{\textbf{Runtime and model size comparison.}
Average inference time and model size for LIVI and baselines. Error bars denote standard deviation across runs.
}
\label{fig:runtime_params_horizontal}
\end{figure}

A central motivation for LIVI is to circumvent the computational cost of Whisper's autoregressive decoding. To quantify efficiency gains, we measure end-to-end latency on $200$ randomly sampled tracks from Discogs-VI, comparing LIVI with the transcription-based pipeline and audio baselines. For all models, we exclude data preprocessing from reported latencies: it is identical across methods, CPU-bound, and can be overlapped with GPU computation in a streaming pipeline, incurring no additional wall-clock cost. We also report model size in terms of trainable parameters (Figure~\ref{fig:runtime_params_horizontal}).

The transcription pipeline requires on average $4.41$\,s per track; LIVI reduces this to $0.446$\,s, a $9.9\times$ end-to-end speed-up. This latency is dominated by the frozen Whisper encoder ($0.429$\,s), while the binary classifier and LIVI projection head together add under $0.02$\,s, underscoring the efficiency of detecting hallucination-prone chunks from representations already computed for embedding. Against the audio baselines ($0.66$--$1.40$\,s), LIVI is $1.5–3.1\times$ faster, with lower latency variance (std.\ $0.12$ vs.\ $1.03$--$2.66$), ensuring more predictable and stable performance. 
LIVI further offers a favorable trade-off between model complexity and accuracy. With $31.9$M trainable parameters ($45.84$\,M including the binary classifier), it is substantially lighter than large systems such as ByteCover2 ($202.3$M) and CLEWS ($196.8$M) while remaining competitive in accuracy, and it surpasses small models such as CQTNet and DViNet, particularly on large benchmarks like Discogs-VI.

At query time, MaxSim reranking over $B_{0.85}(q)$ adds only $0.43$\,ms per query on top of the $17.86$\,ms ball search, owing to the small number of local embeddings per track ($6.0 \pm 2.5$). This same factor bounds the storage overhead: retaining one local embedding per chunk rather than a single global vector increases the index size by approximately $6$ times, from $0.247$\,GB to $1.491$\,GB on Discogs-VI ($80{,}485$ tracks). This remains tractable at this scale, and could be further reduced through quantization, which we leave to future work.

\section{Qualitative Analysis}
\label{sec:qualitative}

In this section, we complement the quantitative results with qualitative insights: we examine the structure of the embedding space, discuss potential limitations of the system, and provide examples illustrating the role of specific components such as the binary classifier and the MaxSim reranker.

\subsection{Per-Genre and Per-Language Breakdown}
\label{sec:breakdown}

We decompose the retrieval performance of our proposed model by musical genre and by language, to determine whether the aggregate metrics of Section~\ref{sec:main_results} reflect uniform quality or conceal weaknesses concentrated in specific subsets.
We conduct this analysis on SHS100k, the only benchmark on which LIVI underperforms audio baselines, making it the most informative setting for probing the limits of a lyrics-centered representation. Its moderate size also renders per-category inspection tractable.
Genre is obtained by matching tracks against our proprietary catalog and retaining the associated editorial tag. Language is estimated through Whisper's native language-identification mechanism. A single decoder forward pass conditioned on the \texttt{<|startoftranscript|>} token yields a probability distribution over Whisper's language tokens, which are averaged across a track's reliable chunks to assign the language with the highest mean probability.

\begin{table}[h!]
\scriptsize
\centering
\begin{subtable}[t]{\linewidth}
\centering
\begin{tabular}{l r ccc}
    \toprule
    \textbf{Genre} & \textbf{\#queries} & \textbf{MR1}$\downarrow$ & \textbf{HR@1}$\uparrow$ & \textbf{MAP@10}$\uparrow$ \\
    \midrule
    Jazz                & 311 & 5.10  & 0.978 & 0.969 \\
    Pop                 & 266 & 10.14 & 0.899 & 0.811 \\
    Brazilian music     & 92  & 1.01  & 0.989 & 0.962 \\
    Rock                & 65  & 11.29 & 0.923 & 0.885 \\
    Latino              & 41  & 1.37  & 0.976 & 0.923 \\
    Vocal jazz          & 35  & 1.00  & 1.000 & 0.969 \\
    Country             & 34  & 1.00  & 1.000 & 0.977 \\
    Classical           & 28  & 1.25  & 0.964 & 0.941 \\
    Folk                & 23  & 25.09 & 0.870 & 0.793 \\
    Alternative         & 20  & 1.00  & 1.000 & 0.886 \\
    \midrule
    Untagged            & 599 & 4.65  & 0.977 & 0.904 \\
    All                 & 1{,}670 & 5.32 & 0.961 & 0.906 \\
    \bottomrule
\end{tabular}
\caption{Per genre.}
\label{tab:shs_genre}
\end{subtable}

\vspace{8pt}

\begin{subtable}[t]{\linewidth}
\centering
\begin{tabular}{l r r ccc}
    \toprule
    \textbf{Language} & \textbf{\#queries} & \textbf{\%X-ling} & \textbf{MR1}$\downarrow$ & \textbf{HR@1}$\uparrow$ & \textbf{MAP@10}$\uparrow$ \\
    \midrule
    English (en)    & 680 & 49.3 & 3.01  & 0.985 & 0.964 \\
    Portuguese (pt) & 200 & 49.4 & 1.34  & 0.980 & 0.959 \\
    Spanish (es)    & 71  & 75.0 & 12.61 & 0.901 & 0.825 \\
    Finnish (fi)    & 28  & 95.3 & 13.25 & 0.679 & 0.345 \\
    Swedish (sv)    & 24  & 68.7 & 28.63 & 0.875 & 0.792 \\
    French (fr)     & 22  & 91.0 & 1.00  & 1.000 & 0.766 \\
    \midrule
    Untagged        & 599 & 48.0 & 4.65  & 0.977 & 0.904 \\
    All           & 1{,}670 & 50.5 & 5.32 & 0.961 & 0.906 \\
    \bottomrule
\end{tabular}
\caption{Per language. \%X-ling: percentage of cross-lingual positive pairs.}
\label{tab:shs_lang}
\end{subtable}

\caption{Per-genre and per-language retrieval performance of the LIVI pipeline on SHS100k, for categories with at least twenty queries.}
\label{tab:shs_breakdown}
\end{table}

Across genres in Table~\ref{tab:shs_genre}, retrieval quality is high and fairly uniform. Performance declines as bucket size grows, but this trend does not fully explain the observed spread. Jazz, the largest category (311 queries), clearly surpasses Pop (266 queries), reaching HR@1 = 0.978 and MAP@10 = 0.969 against Pop's 0.899 and 0.811. A small number of categories (Folk and Rock) show elevated MR1 (25.09 and 11.29) despite high HR@1, indicating that mean rank is inflated by a few deeply mis-ranked queries rather than by a broad degradation; given their limited support (23 and 65 queries), we interpret these tails cautiously.

Table~\ref{tab:shs_lang} shows that performance declines for languages that are underrepresented in the ASR model and text encoder. English and Portuguese, the two largest buckets, achieve the strongest results (HR@1 = 0.985 and 0.980; MAP@10 = 0.964 and 0.959), while performance falls sharply for under-represented languages---especially Finnish, which records the lowest HR@1 (0.679) and MAP@10 (0.345).
For each language, we compute a cross-lingual rate \%X-ling, defined as the proportion of positive pairs in which the two tracks' estimated languages differ. While Finnish and French have comparable cross-lingual rates (95.3\% vs. 91.0\%), retrieval quality differs sharply (MAP@10=0.345 vs. 0.766), suggesting the gap stems from uneven language coverage in Whisper's and the text encoder's pretraining data rather than from cross-lingual matching itself.

A closer, manual analysis reveals two failure modes. The first involves lyric adaptation, where lyrics are mostly rewritten, as in the two English adaptations \textit{"Desafinado (Off Key)"} (translated by Gene Lees) and \textit{"Slightly Out of Tune"} (translated by Jon Hendricks) of the Portuguese original \textit{"Desafinado"} written by Antônio Carlos Jobim and Newton Mendonça. The former preserves the original Portuguese text's literal, self-deprecating humor about vocal ability ("When I try to sing, you say I'm off-key") and its playful, conversational wit ("I took your picture with my trusty Rolleiflex / And now all I have developed is a complex"). Conversely, the latter reimagines the song through romantic metaphors of emotional disconnect ("our song of love is slightly out of tune" and "like the bossa nova, love should swing"). In this case, while the underlying musical identity is preserved, the lexical content is almost entirely rewritten, creating a major obstacle for lyrics-based methods.
The second failure mode concerns covers linked solely by melody, sharing no lyrical content at all. The \textit{"O Tannenbaum"} melody illustrates this challenge, with four texts sharing no lexical content: the German carol itself, the American anthem \textit{"Maryland, My Maryland"} the British anthem \textit{"The Red Flag"} and the Finnish carol \textit{"Oi kuusipuu"}. Yet being genuine covers, none are retrieved within the top 1,500 ranks. Such melody-only covers fall entirely outside the scope of any lyrics-based method, exposing the main limitation of our approach and motivating the use of complementary musical features in future work.

\subsection{Visualisation of the Embedding Space}
\label{sec:visualisation}

\begin{figure*}[!ht]
    \centering
    \begin{minipage}{0.9\linewidth}
        \centering
        \begin{subfigure}[t]{0.4\linewidth}
            \centering
            \includegraphics[height=5.5cm, width=\linewidth, keepaspectratio]{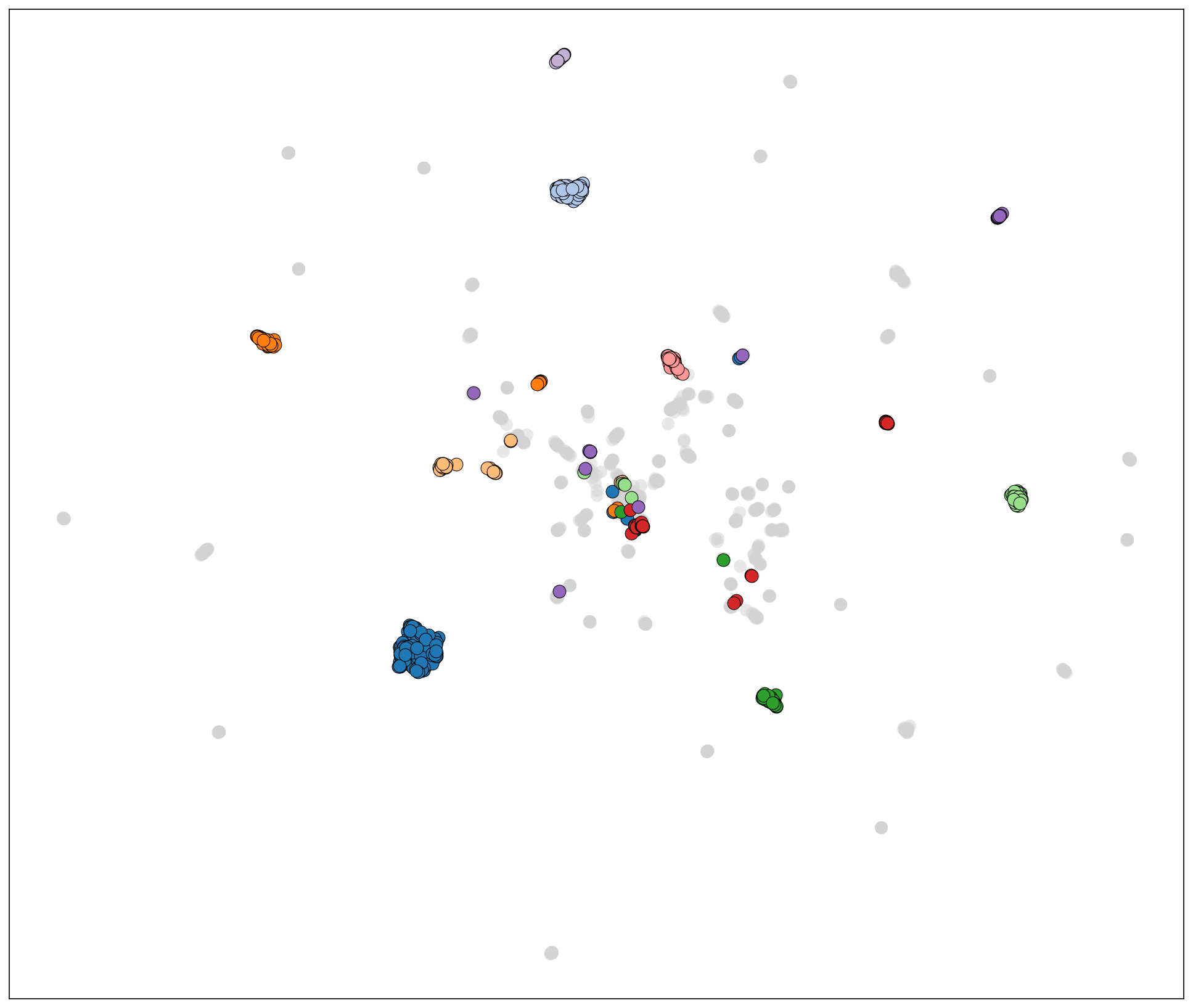}
            \caption{SHS100k---colored by cliques}
        \end{subfigure}
        \hfill
        \begin{subfigure}[t]{0.48\linewidth}
            \centering
            \includegraphics[height=5.5cm, width=\linewidth, keepaspectratio]{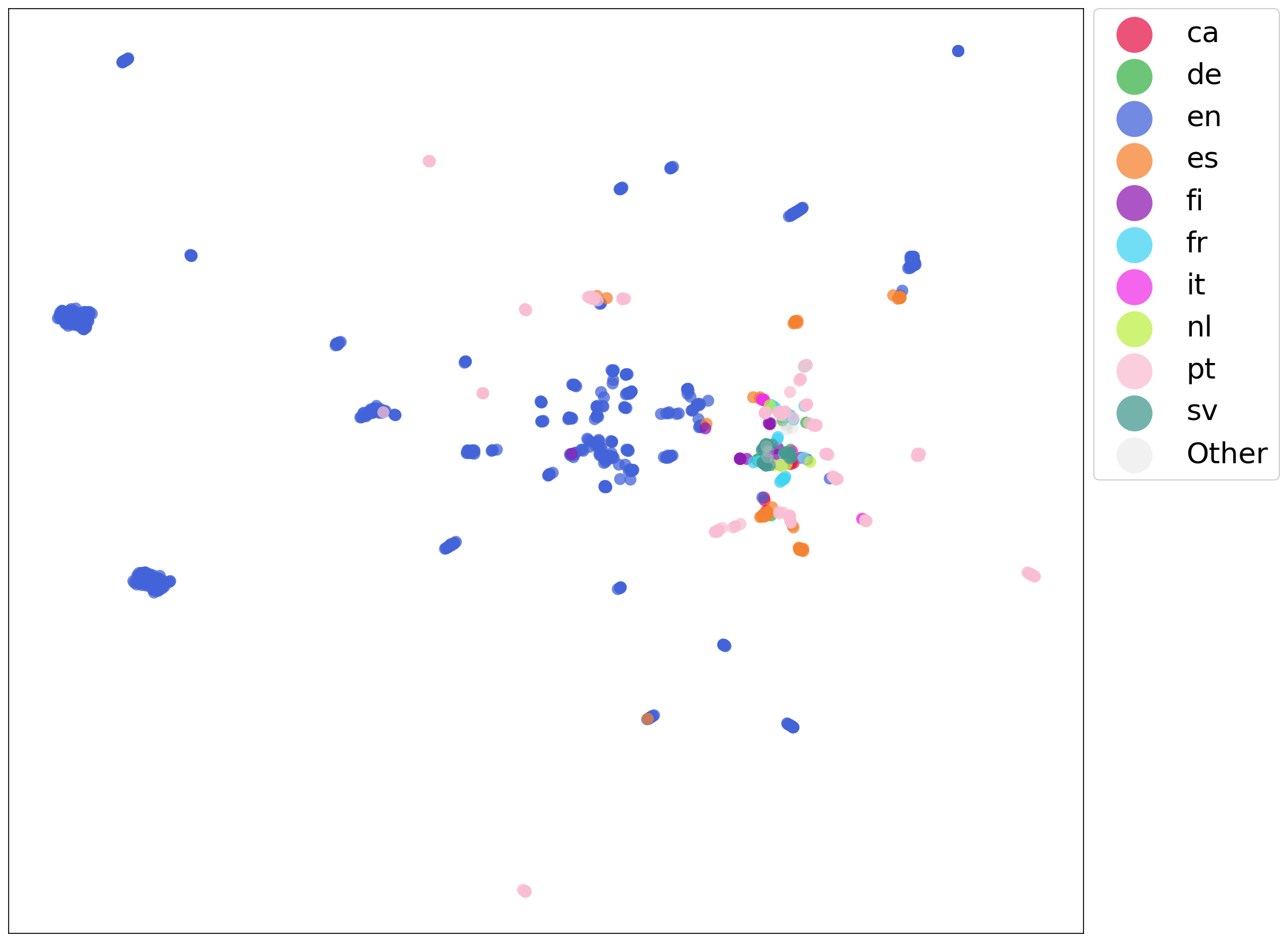}
            \caption{SHS100k---colored by language}
        \end{subfigure}
        \vspace{4pt}
        \begin{subfigure}[t]{0.4\linewidth}
            \centering
            \includegraphics[height=5.5cm, width=\linewidth, keepaspectratio]{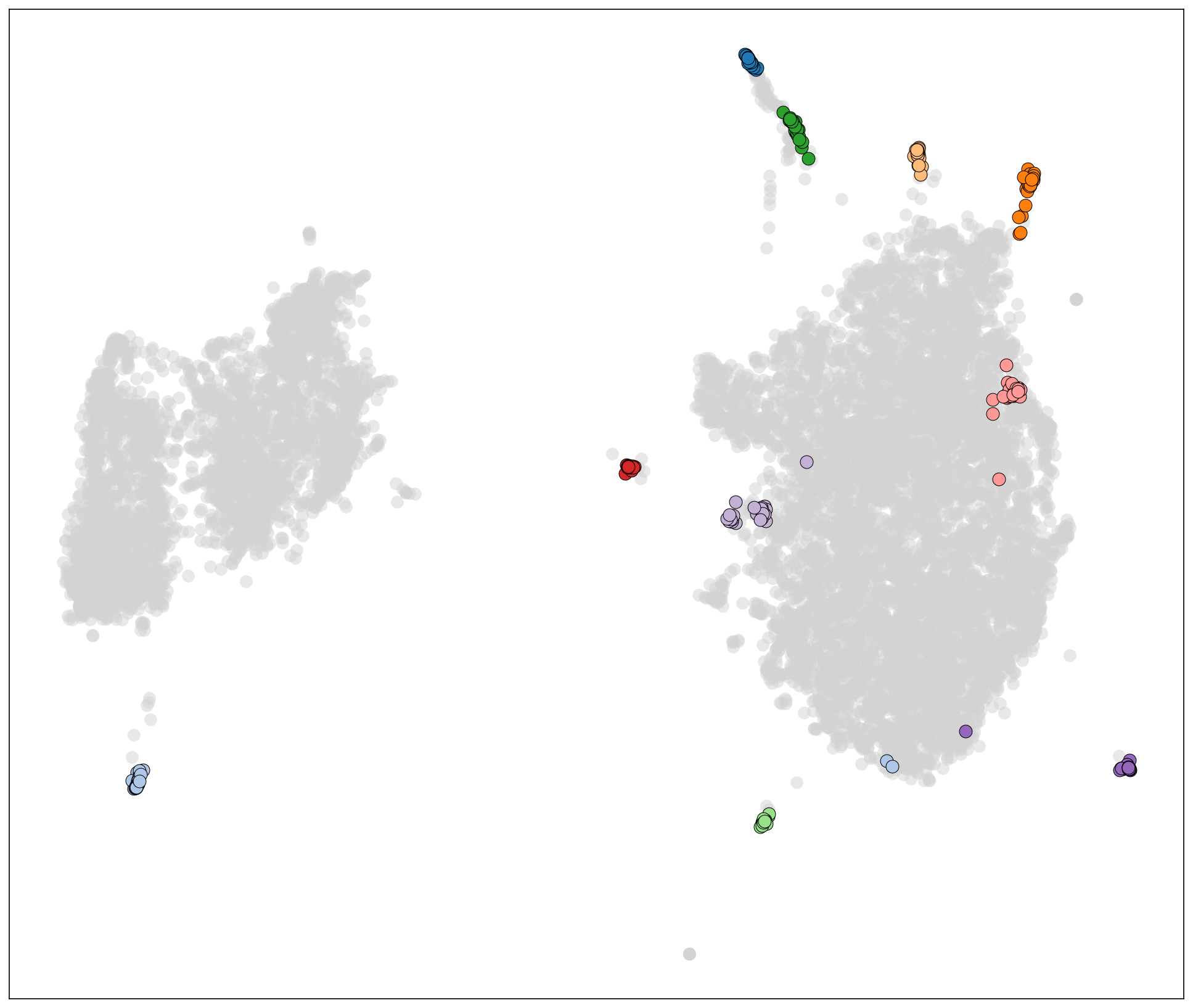}
            \caption{Discogs-VI---colored by cliques}
        \end{subfigure}
        \hfill
        \begin{subfigure}[t]{0.48\linewidth}
            \centering
            \includegraphics[height=5.5cm, width=\linewidth, keepaspectratio]{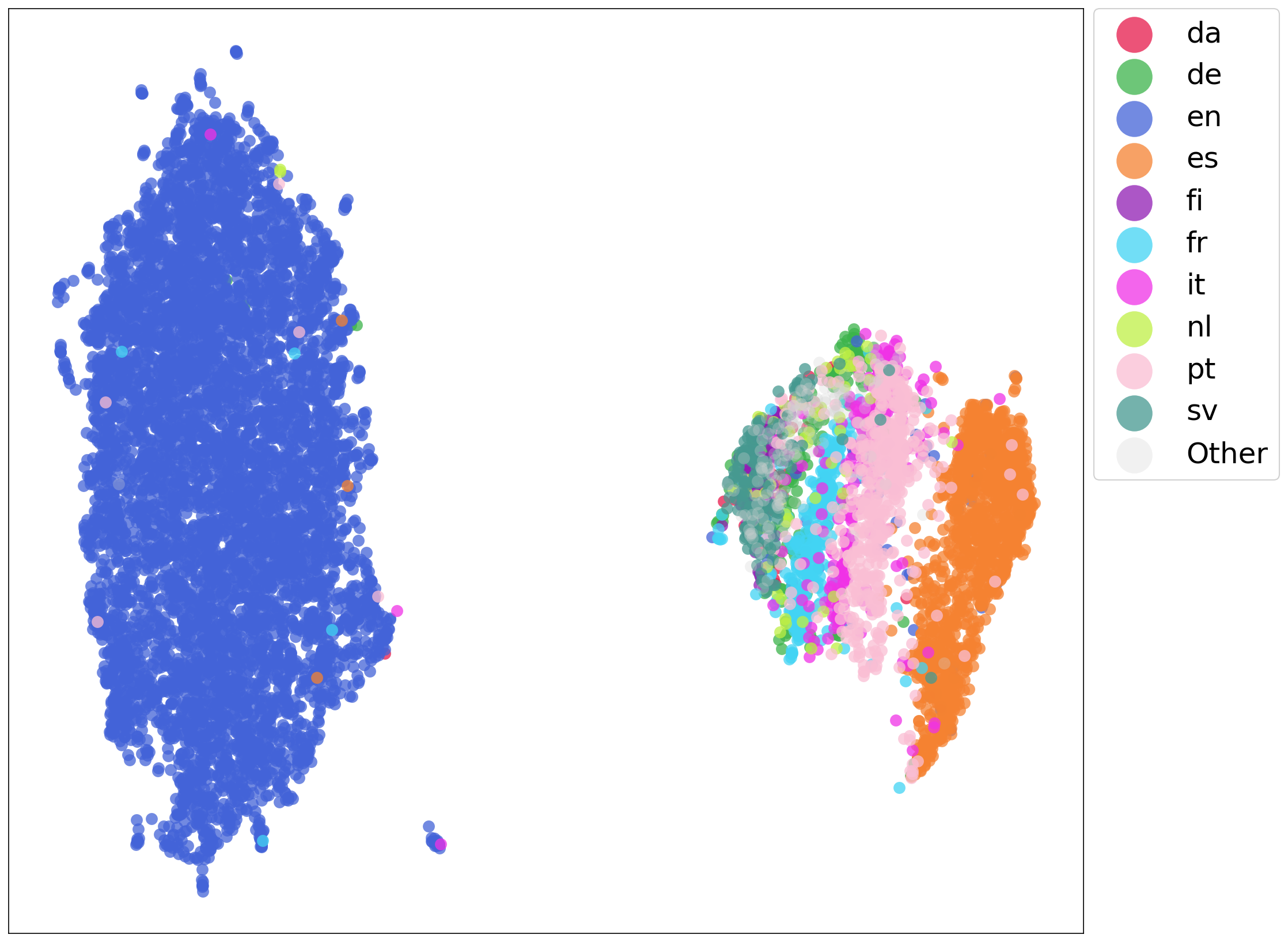}
            \caption{Discogs-VI---colored by language}
        \end{subfigure}
    \end{minipage}
    \caption{UMAP projections of track-level LIVI embeddings on SHS100k and Discogs-VI, showing random cliques (a, c) and language (b, d).}
    \label{fig:umap_overview}
\end{figure*}

We now examine the geometry of the lyrics-informed embedding space through UMAP projections of track-level LIVI embeddings. Figure~\ref{fig:umap_overview} shows randomly sampled version cliques highlighted across SHS100k and Discogs-VI benchmarks. In each case, clique members are drawn into tight, well-separated groups, consistent with the good retrieval performance reported in Table~\ref{tab:vi_main}.
Coloring the same projections by language interestingly illustrates a potential bias in the method. On Discogs-VI (Figure~\ref{fig:umap_overview}.d), the space separates into two regions: a large region dominated by English tracks, and a smaller, more heterogeneous region grouping the remaining languages. Rather than interleaving languages according to shared lyrical meaning, as a fully language-agnostic space would, the geometry retains a strong residual organization by language, with English occupying a distinct and self-contained region.
This residual language structure likely hints a cross-lingual limitation of our model, likely cascading from uneven multilingual performance in both the Whisper encoder and the text embedding model. However, the absence of such a phenomenon for SHS100k data in Figure~\ref{fig:umap_overview}.b, with a much more skewed and narrower language distribution, provides the alternative explanation that the bias likely originates in the data itself. This might come from non-English language matching limitations in Discogs-VI creation, as acknowledged by the authors \citep{araz2024discogsvimusicalversionidentification}, or from the general observation that the Discogs collection is strongly biased towards English releases in the first place \citep{bogdanov2017quantifying}). 

\subsection{Impact of Removing Hallucination-Prone Chunks}
\label{sec:umap}

\begin{figure}[!t]
    \centering
    \begin{minipage}{0.8\linewidth}
        \centering
        \begin{subfigure}[t]{0.48\linewidth}
            \centering
            \includegraphics[width=\linewidth]{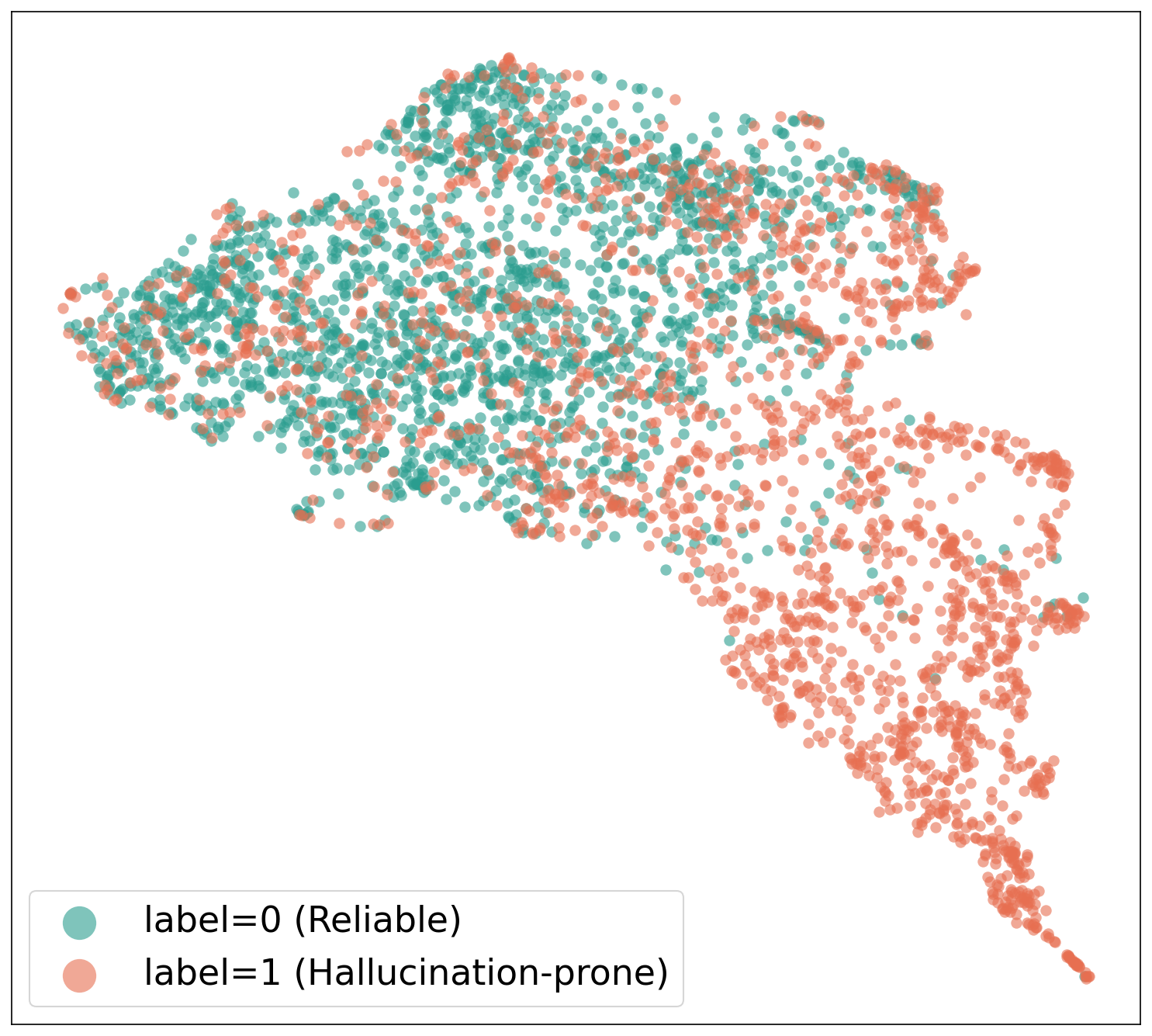}
            \caption{Before filtering: reliable and hallucination-prone chunks.}
            \label{fig:umap_labels_all}
        \end{subfigure}
        \hfill
        \begin{subfigure}[t]{0.48\linewidth}
            \centering
            \includegraphics[width=\linewidth]{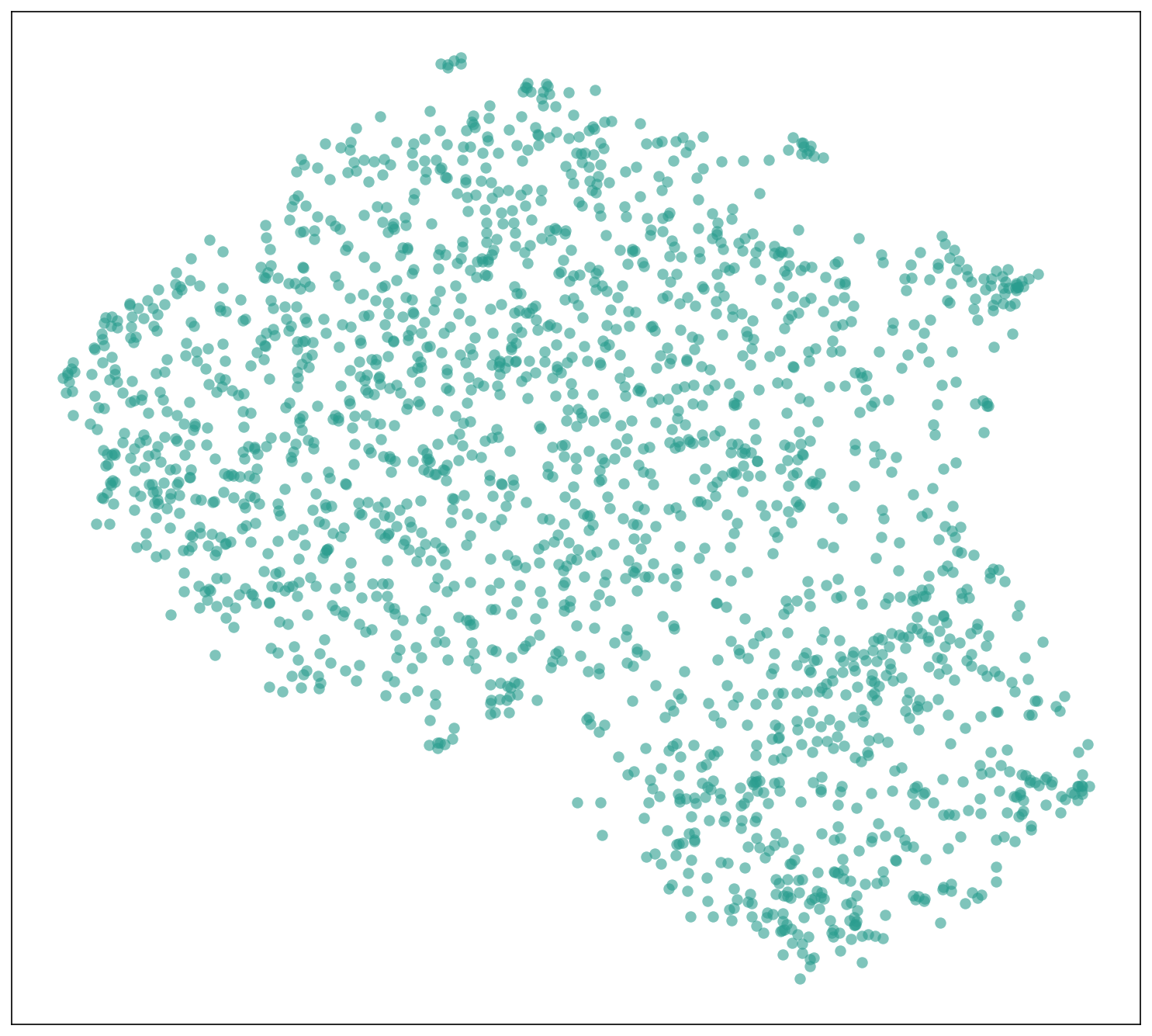}
            \caption{After filtering: reliable chunks only.}
            \label{fig:umap_labels_reliable}
        \end{subfigure}
    \end{minipage}
    \caption{UMAP projections of 4,000 chunk-level LIVI embeddings, evenly split
    between reliable and hallucination-prone predictions from the binary
    classifier.}
    \label{fig:umap_labels}
\end{figure}

\begin{table}[h!]
\centering
\small
\begin{tabular}{p{0.03\linewidth} p{0.17\linewidth} p{0.16\linewidth} p{0.12\linewidth} p{0.38\linewidth}}
\toprule
& \textbf{Title} & \textbf{Artist} & \textbf{Genre} & \textbf{Whisper transcription} \\
\midrule
$(i)$ & Sugar Sugar  & Birds 'N Brass & Jazz & \textit{``We'll be right back.''} \\
\addlinespace
$(ii)$ & Proud Mary & Señor Soul  & Jazz & \textit{``Go! Go! Go! Go! Yeah! Go! Go! Go! Go! Go! Go! Go! you Go! Go! Go! Go! Go! Go! Go! Go! Go! Go! Go! Go! Go! Go! Go! Now, here! Yeah! Power! Roll on, baby Yeah Yeah Roll down down''} \\
\addlinespace
$(iii)$ & Smile & Carolina Eyck & Scat singing & \textit{``I'm going to die. Sabare Sabare So What are our own Oh Oh''} \\
\addlinespace
$(iv)$ & Ave Maria & Fischer Chöre, Gotthilf Fischer & Opera & \textit{``The Lord is the Son of God.''} \\
\addlinespace
$(v)$ & Promise & Coffin Break & Rock & \textit{``We'll be right back.''} \\
\addlinespace
$(vi)$ & Somewhere Over the Rainbow & Hank Roberts & Experimental & \textit{``What? Just what? Just what? Just what? I'm I'm I'm I'm I'm I'm I'm I'm I'm I'm I'm I'm I'm I'm To play...''} \\
\addlinespace
$(vii)$ & Yesterday & Danielle Licari & Pop & \textit{``I'm I'm I'm I'm I'm I'm''} \\
\bottomrule
\end{tabular}
\caption{Whisper transcriptions of tracks with all chunks identified as hallucination-prone by the binary classifier despite passing the proprietary vocalness gate.  Examples are reported at the track level.
}
\label{tab:new_rejections}
\end{table}

We turn next to a visual analysis of how hallucination-prone chunks affect the embedding space. Figure~\ref{fig:umap_labels} shows UMAP projections of a balanced sample of 4,000 chunk-level LIVI embeddings, evenly split between chunks predicted as reliable and hallucination-prone by the binary classifier. Before filtering, hallucination-prone chunks occupy a distinct, elongated region, mostly separated from the main cloud of reliable chunks. This reflects the Bag of Hallucinations artifacts described in Section~\ref{sec:weak_labels}, whose recurring phrases (e.g.\ \textit{``thank you''} or \textit{``guitar solo''}) collapse musically unrelated chunks into the same neighbourhood of the space. Retaining only the chunks predicted as reliable removes this region: the remaining embeddings form a relatively uniform cloud with no residual trace of the hallucination-driven cluster, consistent with the binary classifier successfully isolating the chunks responsible for polluting the embedding space.

To ground this pattern in concrete examples, Table~\ref{tab:new_rejections} shows representative Whisper transcriptions of tracks identified as hallucination-prone by the binary classifier despite passing the proprietary vocalness gate (Section~\ref{sec:weak_labels}). The most affected genres are jazz and opera, where atypical vocal styles fall outside of Whisper's training distribution and elicit hallucinations rather than genuine lyric transcriptions. Typical hallucination-prone chunks include frequently repeated sentences $(i), (v)$, in-context hallucinations $(iv)$, repeated words and patterns likely linked to auto-regressive decoders $(ii), (vi), (vii)$, and mondegreen-like misrecognitions $(iii)$.

\subsection{MaxSim: Local vs.\ Global Similarity}
\label{sec:maxsim_interp}

\begin{figure}[!b]
    \centering
    \begin{subfigure}[t]{0.8\linewidth}
        \centering
        \includegraphics[width=\linewidth]{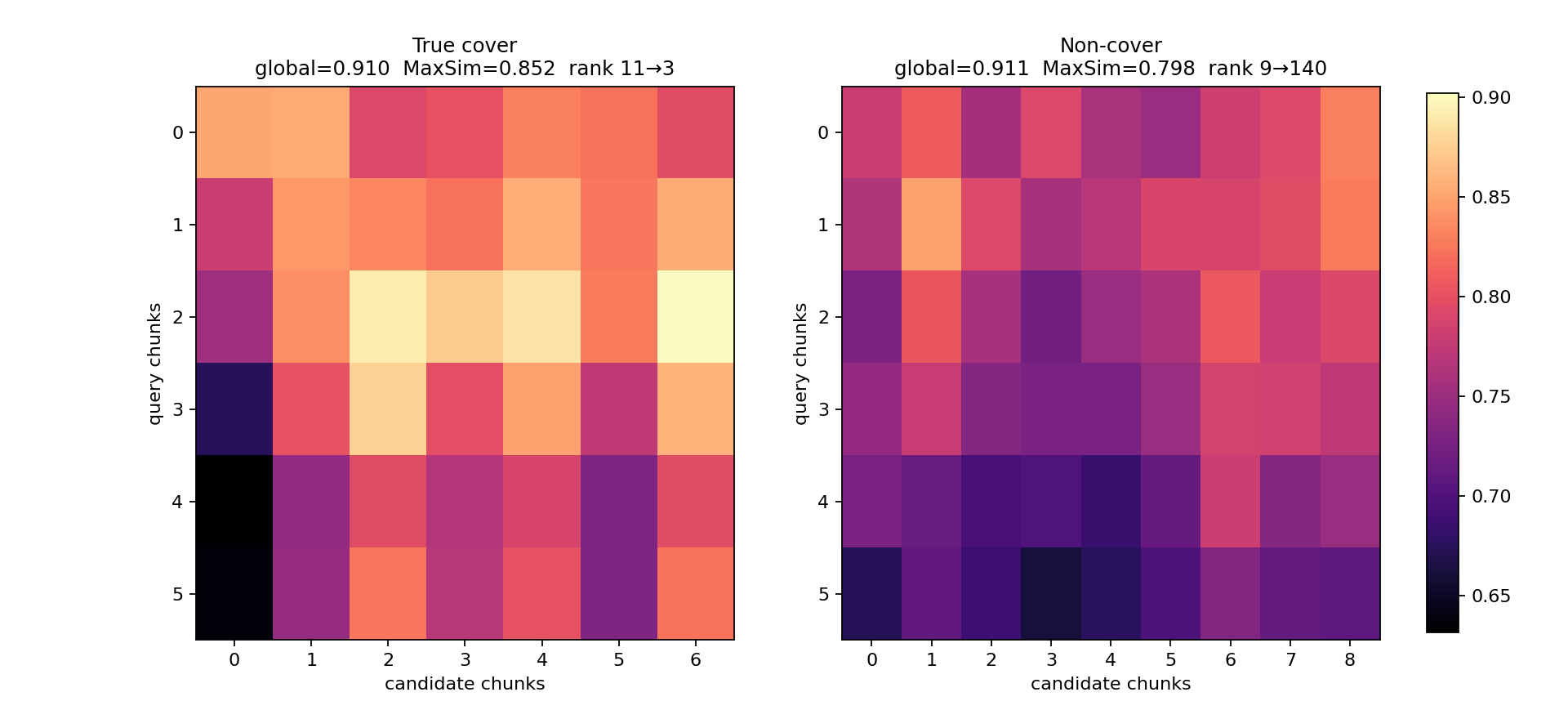}
    \end{subfigure}
    \\[4pt]
    \begin{subfigure}[t]{0.8\linewidth}
        \centering
        \includegraphics[width=\linewidth]{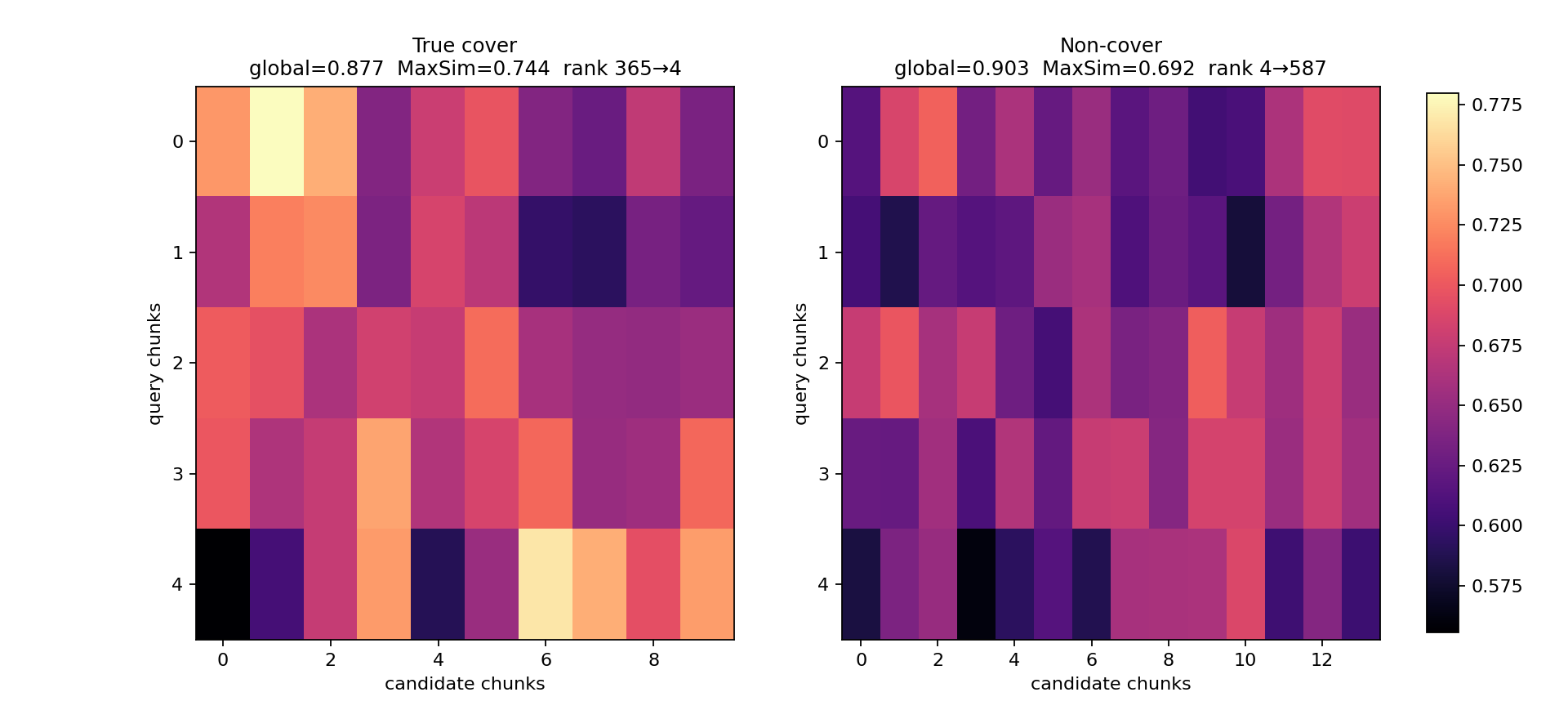}

    \end{subfigure}

    \caption{Locality matrices $\mathbf{S}$ for two query tracks, each
    paired with a true cover (left) and a non-cover (right).
    Each cell $(i,j)$ reports the cosine similarity between query chunk $i$
    and candidate chunk $j$; brighter colours indicate higher similarity.
    }
    \label{fig:locality_matrices}
\end{figure}

In Section~\ref{sec:inference}, we introduced the MaxSim operator to confer robustness to variations characteristic of covers that averaging into a single global vector would dilute, such as structural reordering, partial overlap, and differing track lengths. In Figure~\ref{fig:locality_matrices}, we illustrate the MaxSim calculation relevance by displaying the locality matrix $\mathbf{S}\in\mathbb{R}^{N_q\times N_c}$ between queries $a_i^{(q)}$ and their matching candidates $a_j^{(c)}$ as follows:
$S_{ij}=\cos(a_i^{(q)},a_j^{(c)})$.

The figure illustrates how MaxSim disambiguates true covers from non-covers, across a Metal query (top) and a Rap query (bottom). In each case, the true cover (left) and the non-cover (right) yield comparable global cosine scores, yet their locality matrices differ substantially. 
For the true cover, the matrices exhibit one or more concentrated bright regions reflecting a shared distinctive passage. MaxSim recovers the strong localized match and promotes it, both for the Metal query ($\text{global}=0.910$, $\text{MaxSim}=0.852$, rank $11\rightarrow3$) and the Rap query ($\text{global}=0.877$, $\text{MaxSim}=0.744$, rank $365\rightarrow4$).
Conversely, the non-cover matrices are broadly warm but lack any sharp peak: the candidate shares vocabulary with the query without reproducing a common passage, inflating the global cosine while leaving no strongly matching chunk pair. MaxSim demotes the false positive accordingly, with rank falling from $9$ to $140$ for the Metal query ($\text{global}=0.911$, $\text{MaxSim}=0.798$) and from $4$ to $587$ for the Rap query ($\text{global}=0.903$, $\text{MaxSim}=0.692$).
Together, the two cases illustrate how local alignment disambiguates what global similarity cannot.

\section{Conclusion}
\label{sec:conclusion}

This work introduced LIVI, an approach for Music Cover Retrieval that balances retrieval accuracy with computational efficiency. Our starting point is a lyrics-informed pipeline that transcribes audio with an encoder–decoder ASR model and embeds the resulting text with a multilingual model fine-tuned for semantic similarity, yielding a space where semantically similar lyrics cluster closely. Its key drawback is the ASR decoder's heavy computational overhead, a bottleneck LIVI removes by training an audio encoder to map ASR encoder states directly into this embedding space, discarding the decoder entirely.
With no task-specific fine-tuning and a relatively simple architecture, it stands in contrast to the prevailing trend toward increasingly complex and resource-intensive models. Yet, when tracks have lyrics, LIVI achieves performance on par with these systems, demonstrating that musically informed inputs, rather than architectural escalation, can deliver state-of-the-art results while remaining computationally efficient at both training and inference.

Several limitations of the proposed approach should be acknowledged. First, the text embedding model is used off-the-shelf and is not fine-tuned specifically for the retrieval task, leaving room for further performance gains. 
Second, LIVI is inherently restricted to musical content with sufficient vocal material. Purely instrumental tracks or songs with minimal vocals are therefore excluded, limiting the universality of the method and making its applicability dependent on the availability and quality of lyrics information. Incorporating complementary harmonic features could help mitigate this limitation, particularly for cover versions that preserve melodic structure while substantially altering lyrics or falling outside the scope of the current model. 
Third, our qualitative analysis reveals that the learned space retains a residual organisation by language, the geometric signature of a cross-lingual weakness that depresses retrieval for under-represented languages. This limitation is likely inherited from the two large multilingual models composing our pipeline---the ASR and text encoders---and reflects a broader, unresolved challenge in multilingual representation learning rather than one specific to LIVI. Better-balanced multilingual supervision or explicit cross-lingual alignment could nonetheless reduce it.
Finally, at catalog scale, the local-embedding index grows roughly sixfold compared to storing a single global vector per track; quantizing these embeddings would help curb this storage overhead while preserving the benefits of reranking. We leave this direction to future work.

More broadly, LIVI illustrates that grounding retrieval in a stable invariant such as the lyrics can rival architectural complexity, and we expect this principle to extend to other information retrieval tasks where such invariants can be identified and supervised at scale. Furthermore, the LIVI embedding space has been validated solely on Music Cover Retrieval in this work. We nonetheless believe its semantic grounding in lyrical content holds promise for other downstream tasks that depend on textual or thematic similarity, such as mood classification, topic identification, music recommendation, and cross-modal audio-to-text and text-to-audio retrieval. We highlight these as avenues worth exploring in future work.

\acks{
This work was carried out at Deezer Research and funded by Deezer. All authors are or have been employed by Deezer, which operates a commercial music-streaming service for which music cover retrieval is of direct commercial relevance; the authors declare no other competing interests.
}

\appendix

\section{Bag of Hallucinations}

 \begin{center}
\small
\begin{longtable}{l r r}
\caption{Most frequent hallucinated phrases identified in the Bag of Hallucinations
(BoH), with their occurrence counts and percentages across the $1{,}023{,}917$ non-vocal (instrumental) chunks in the dataset.}
\label{app:boh} \\
\toprule
\textbf{Hallucination} & \textbf{Occurrences} & \textbf{Percentage} \\
\midrule
\endfirsthead

\multicolumn{3}{c}{\tablename\ \thetable{} -- continued from previous page} \\
\toprule
\textbf{Hallucination} & \textbf{Occurrences} & \textbf{Percentage} \\
\midrule
\endhead

\bottomrule
\multicolumn{3}{r}{\textit{Continued on next page}} \\
\endfoot

\bottomrule
\endlastfoot

thank you & 567{,}254 & 55.68 \\
guitar solo & 132{,}703 & 13.03 \\
the end & 47{,}800 & 4.69 \\
music & 17{,}134 & 1.68 \\
im out & 8{,}406 & 0.83 \\
well be right back & 5{,}714 & 0.56\\
piano plays softly & 5{,}473 & 0.54 \\
bye & 4{,}252 & 0.42 \\
oh my god & 3{,}927 & 0.39\\
oh & 3{,}196 & 0.31 \\
orchestra plays & 3{,}070 & 0.30 \\
outro music & 2{,}689 & 0.26 \\
piano plays & 2{,}391 & 0.23 \\
música & 2{,}377 & 0.23 \\
¡suscríbete al canal & 2{,}051 & 0.20\\
well see you next time & 1{,}994 & 0.20 \\
music you & 1{,}659 & 0.16\\
yeah & 1{,}595 & 0.16 \\
music thank you & 1{,}534 & 0.15 \\
ill see you next time & 1{,}218 & 0.12\\
zither harp & 1{,}128 & 0.11 \\
the you & 1{,}006 & 0.10\\
the thank you & 826 & 0.08\\
oh yeah & 792 & 0.08 \\
uh & 722 & 0.07 \\
e aí & 711 & 0.07 \\
to be continued & 692 & 0.07 \\
altyazı mk & 659 & 0.06 \\
hey & 623 & 0.06\\
organ plays & 609 & 0.06\\
all right & 603 & 0.06 \\
piano music & 585 & 0.06\\

\end{longtable}
\end{center}

\vskip 0.2in

\bibliography{sample}

\end{document}